\documentclass[apj]{emulateapj}
\submitted{}
\usepackage{apjfonts}

\newcommand{\beq}{\begin{equation}}
\newcommand{\eeq}{\end{equation}}
\newcommand{\beqr}{\begin{eqnarray} \nonumber}
\newcommand{\eeqr}{\end{eqnarray}}

\newcommand{\unit}[1]{\hat{\mathbf{#1}}}

\newcommand{\cm}{\mbox{ cm}}
\newcommand{\sr}{\mbox{ sr}}
\newcommand{\se}{\mbox{ s}}
\newcommand{\yr}{\mbox{ yr}}
\newcommand{\erg}{\mbox{ erg}}
\newcommand{\Hz}{\mbox{ Hz}}
\newcommand{\kHz}{\mbox{ kHz}}
\newcommand{\MHz}{\mbox{ MHz}}
\newcommand{\GHz}{\mbox{ GHz}}

\newcommand{\km}{\mbox{ km}}
\newcommand{\pc}{\mbox{ pc}}
\newcommand{\kpc}{\mbox{ kpc}}
\newcommand{\Mpc}{\mbox{ Mpc}}
\newcommand{\eV}{\mbox{ eV}}
\newcommand{\keV}{\mbox{ keV}}

\newcommand{\TeV}{\mbox{ TeV}}
\newcommand{\mK}{\mbox{ mK}}

\newcommand{\K}{\mbox{ K}}
\newcommand{\de}{^{\circ}}

\newcommand{\muG}{\mbox{ $\mu$G}}
\newcommand{\Jy}{\mbox{ Jy}}
\newcommand{\mJy}{\mbox{ mJy}}
\newcommand{\muJy}{\,\mu\mbox{Jy}}
\newcommand{\fin}{\mbox{ .}}
\newcommand{\coma}{\mbox{ ,}}

\newcommand{\gama}{$\gamma$}
\newcommand{\HI}{$\mbox{H \small I}$ }
\newcommand{\HII}{$\mbox{H \small II}$ }

\newcommand{\IUnits}{\erg\se^{-1}\cm^{-2}\sr^{-1}\Hz^{-1} }
\newcommand{\JUnits}{\erg\se^{-1}\cm^{-3}\sr^{-1}\Hz^{-1} }
\newcommand{\ICUnits}{\keV\se^{-1}\cm^{-2}\sr^{-1} }
\newcommand{\synUnits}{\erg\se^{-1}\cm^{-2}\sr^{-1} }
\newcommand{\LCDM}{$\Lambda$CDM }
\newcommand{\Lya}{Ly$\alpha$ }

\begin{document}

\title{Imprint of Intergalactic Shocks on the Radio Sky}

\author{Uri Keshet\altaffilmark{1}, Eli Waxman\altaffilmark{1} 
and Abraham Loeb\altaffilmark{2,3}}
\altaffiltext{1}{Physics Faculty, Weizmann Institute, Rehovot 76100, Israel; keshet@wicc.weizmann.ac.il, waxman@wicc.weizmann.ac.il}
\altaffiltext{2}{Astronomy Department, 
Harvard University, 60 Garden Street, Cambridge, 
MA 02138, USA}
\altaffiltext{3}{Einstein Minerva Center, Physics Faculty,
Weizmann Institute of Science}



\begin{abstract}
Strong intergalactic shocks are a natural consequence of structure
formation in the universe. These shocks are expected to deposit large
fractions of their thermal energy in relativistic electrons ($\xi_e\simeq
0.05$ according to supernova remnant observations) and magnetic fields
($\xi_B\simeq 0.01$ according to cluster halo observations). We calculate
the synchrotron emission from such shocks using an analytical model,
calibrated and verified based on a hydrodynamical \LCDM simulation. The
resulting signal composes a large fraction (up to a few $10\%$) of the
extragalactic radio background below $500 \MHz$. The associated angular
fluctuations, e.g. $\delta T_l \ga 260 (\xi_e\xi_B/5\times 10^{-4})
(\nu/100\MHz)^{-3}\K$ for multipoles $400\la l\la 2000$, dominate the radio
sky for frequencies $\nu \la 10 \GHz$ and angular scales $1\arcmin\la\theta
< 1\de$ (after a modest removal of discrete sources), provided that $\xi_e
\,\xi_B\ga 3\times 10^{-4}$. The fluctuating signal is most pronounced for
$\nu \la 500 \MHz$, dominating the sky there even for $\xi_e\,\xi_B =
5\times 10^{-5}$. The signal will be easily observable by next generation
telescopes such as LOFAR and SKA, and is marginally observable with
present-day radio telescopes. The signal could also be identified through a
cross-correlation with tracers of large-scale structure (such as \gama-ray
emission from intergalactic shocks), possibly even in existing $\la 10\GHz$
CMB anisotropy maps and high resolution $\sim 1\GHz$ radio
surveys. Detection of the signal will provide the first identification of
intergalactic shocks and of the warm-hot intergalactic medium (believed to
contain most of the baryons in the low-redshift universe), and gauge the
unknown strength of the intergalactic magnetic field. We analyze existing
observations of the diffuse radio background below $500\MHz$, and show that
they are well fit by a simple, double-disk Galactic model, precluding a
direct identification of the diffuse extragalactic radio
background. Modelling the frequency-dependent anisotropy pattern observed
at very low ($1$--$10\MHz$) frequencies can be used to disentangle the
distributions of Galactic cosmic-rays, ionized gas, and magnetic
fields. Space missions such as the {\it Astronomical Low Frequency Array}
(ALFA) will thus provide an important insight into the structure and
composition of our Galaxy.
\end{abstract}



\section{Introduction}
\label{sec:Introduction}
\indent

The gravitational formation of structure in the universe inevitably produced strong, collisionless shocks in the intergalactic medium (IGM), owing to the convergence of large-scale flows. In these shocks, electrons are expected to be Fermi accelerated up to highly relativistic ($\ga 10 \TeV$) energies, limited by inverse-Compton cooling off cosmic microwave background (CMB) photons \cite{LoebWaxman2000}. The resulting \gama-ray emission is expected to trace the large-scale structure of the universe. Rich galaxy clusters, characterized by strong, high velocity accretion shocks, should be detected by future \gama-ray missions as bright \gama-ray sources \cite{LoebWaxman2000, Totani00, WaxmanLoeb2000} in the form of accretion rings with bright spots at their intersections with galaxy filaments \cite{Keshet03}. Recently, a possible association of $\gamma$-ray radiation (as measured by the \emph{Energetic Gamma-Ray Experiment Telescope}, EGRET) with the locations of Abell clusters was identified at a $3\sigma$ confidence level \cite[][but see also Reimer et al. 2003]{Scharf2002}. The integrated \gama-ray background resulting from strong intergalactic shocks was calculated using hydrodynamical cosmological simulations \cite{Keshet03,Miniati02}, at the level of $\epsilon^2 dJ/d\epsilon \la 0.15 \keV \se^{-1} \cm^{-2} \sr^{-1}$. This signal constitutes $\la 15\%$ of the widely accepted estimates for the flux of the extragalactic \gama-ray background \cite[EGRB, see e.g.] [] {Sreekumar98,Strong03}, although a recent analysis \cite{Keshet04a} suggests that the EGRB flux is lower than previously thought, at least by a factor of 2. The \gama-ray background from weak merger shocks was estimated to be a factor of $>10$ weaker than the background from intergalactic accretion shocks \cite{Gabici03}. 

In addition to the inverse-Compton emission from intergalactic shocks,
synchrotron radiation should also be emitted by the relativistic electrons, as they gyrate in the shock-induced magnetic fields. The resulting radio signature is expected to trace the structure of the universe at low redshifts ($z\la1$), and to be dominated by rich, young galaxy clusters. Indeed, extended radio emission with no optical counterpart is observed in about $10\%$ of the rich galaxy clusters \cite[radio halos and radio relics, see ] [] {Giovannini99}, and in more than a third of the young, massive clusters \cite[with X-ray luminosity $L_X>10^{45} \erg \se^{-1}$, see][]{Feretti03}. The radio emission has been identified as synchrotron radiation from relativistic electrons, but there are different models for the origin of the electrons and the magnetic fields involved \cite[for a recent discussion, see][]{Bagchi03}. 
En\ss{}lin et al. (1998) have used observations of nine radio relics to suggest an association between these sources and structure formation shocks, focusing on the possibility that the relics are revived fossil radio cocoons originating from nearby radio galaxies, re-energized by diffusive shock acceleration. It is interesting to note that recently, large-scale diffuse radio emission was discovered around a filament of galaxies, possibly tracing an accretion shock on this scale \cite{Bagchi_etal02}.

Waxman \& Loeb (2000) have proposed a simple model, which allows one to
estimate both the radio and the \gama-ray signatures of intergalactic shocks, produced by electrons accelerated from the inflowing plasma. Their model allows one to estimate the radio and the \gama-ray backgrounds and their anisotropy characteristics, as well as the signature of individual clusters. The model uses dimensional analysis arguments to estimate the properties of the virialization accretion shock of a halo, as a function of redshift $z$ and halo mass $M$. Halo abundance estimates at different redshifts [such as the Press \& Schechter (1974) halo mass function], may then be used to calculate various observables. The Waxman \& Loeb model approximates the strong accretion shocks as being spherically symmetric, and neglects weak merger shocks. A fraction $\xi_e \simeq 0.05$ of the shock thermal energy is assumed to be carried by relativistic electrons, based on supernovae remnant (SNR) observations \cite[for a discussion, see] [] {Keshet03}. Observations of $\ga 0.1 \muG$ magnetic fields in cluster halos \cite{Kim89, Fusco-Femiano99, Rephaeli99} require that a fraction $\xi_B\simeq 0.01$ of the shock thermal energy be transferred into downstream magnetic fields. With these assumptions, Waxman \& Loeb (2000) found that strong fluctuations in the predicted radio signal seriously contaminate CMB anisotropy measurements at sub-degree angular scales and frequencies below $10\GHz$.

Identification of radio or \gama-ray emission from intergalactic shocks
holds a great promise for advancing current knowledge of shock formation in
the IGM. It should provide the first direct evidence for such shocks,
revealing the underlying large-scale cosmological flows. When combined with
\gama-ray detection, the radio signal will provide a direct measure of the
unknown magnetic fields in the IGM, possibly shedding light on the processes leading to IGM magnetization. Emission from large-scale shocks traces the undetected warm-hot IGM at temperatures $10^5\K \la T\la 10^7 \K$ that is
believed to contain most of the baryons in the low redshift universe
\cite[see e.g.][]{Davea01}. Moreover, the signal may be used to study
non-thermal physical processes in the intergalactic environment, such as
Fermi acceleration in low density shocks.

In this paper we explore the radio signature of intergalactic shocks and
examine its observational consequences. Our results derive from a generalized version of the model of Waxman \& Loeb (2000), adapted for a \LCDM universe,
modified to incorporate non-spherical accretion shocks, and calibrated
using the global features of a simulated \LCDM universe in a hydrodynamical cosmological simulation. We perform an analysis of the radio sky in order to assess the feasibility of observing the predicted signal; in this context, we review various known foreground and background radio signals, and analyze observations at low ($\nu < 500\MHz$) frequencies, where the diffuse extragalactic background is most pronounced with respect to the Galactic foreground. Our results have implications for existing high resolution radio telescopes (e.g. the Very Large Array\footnote{see http://www.vla.nrao.edu/astro/guides/vlas}); for next generation ground-based radio telescopes such as the LOw Frequency Array (LOFAR\footnote{see http://www.lofar.org}) and the Square Kilometer Array (SKA\footnote{see http://www.skatelescope.org}); and for future space-borne telescopes such as the Astronomical Low Frequency Array (ALFA\footnote{see http://sgra.jpl.nasa.gov/html\_dj/ALFA.html}).
 
In \S\ref{sec:IGM_model} we calculate the radio signal from intergalactic
shocks. We start by generalizing the model of Waxman \& Loeb (2000, illustrated in Figure \ref{fig:PS_illustration}), and adapting it for a \LCDM universe. The non-uniformity of the thermal injection rate along the shock front is incorporated into the model by modifying the modelled shock front area. The free parameters of the model are calibrated using \emph{global} features (such as the average baryon temperature) of a simulated \LCDM universe, according to a previously analyzed hydrodynamical cosmological simulation \cite{Springel2001}. The predictions of the calibrated model are then shown to agree with the results of the cosmological simulation, regarding the radio emission (see Figures \ref{fig:calculated_spectrum}, \ref{fig:calculated_correlation} and \ref{fig:synch_Cl}) and its \gama-ray counterpart \cite[see][]{Keshet03}. We estimate the energy fractions $\xi_e$ and $\xi_B$ using SNR and cluster halo observations, and evaluate the uncertainty of these parameters. A qualitative agreement is shown to exist between observations of galaxy-cluster radio halos, and model predictions (see Figure \ref{fig:clusters}). 

In \S\ref{sec:feasibility} we examine the observational consequences of the radio signal for present and future radio telescopes. We assess the contribution of various foreground and background signals to the brightness of the radio sky (see Figure \ref{fig:LFRB_sky}) and to the angular power spectrum (APS, see Figure \ref{fig:Sky_Cl}). In particular, we examine the synchrotron foreground from our Galaxy, and the contamination from discrete radio sources. We also calculate the point-source removal threshold required so that the fluctuations in the intergalactic shock signal dominate the angular power spectrum on small angular scales. Possible methods by which the signal may be identified using present and future observations are discussed. Other extragalactic radio signals, namely bremsstrahlung emission from \Lya clouds and 21 cm tomography, are also reviewed.

In \S\ref{sec:LFRB} we analyze the diffuse low frequency ($\nu <500 \MHz$)
radio background (LFRB). After highlighting important observational features (see Figure \ref{fig:LFRB_raw}), we present a model for the Galactic foreground that allows us to (i) try to separate between the Galactic foreground and the extragalactic background; (ii) examine if a simple Galactic model can account for the observed Galactic foreground; and (iii) demonstrate the importance of observations in the frequency range $1\MHz \la \nu \la 10\MHz$, where absorption in our Galaxy is significant. We show that existing observational data is consistent with a simple double-disk Galactic model (see Figure \ref{fig:LFRB_model}). The implied lack of direct evidence for a diffuse extragalactic radio background (DERB) is discussed, given earlier models for the LFRB and unsubstantiated claims for direct identifications of the extragalactic component. Finally, we show how future observations at very low ($1-10\MHz$) frequencies, e.g. by the ALFA, may be used to disentangle the distributions of Galactic cosmic-rays, magnetic fields and ionized gas. 

Finally, \S\ref{sec:discussion} summarizes our results and addresses their
potential implications. We discuss the conditions under which emission from
intergalactic shocks could be identified, taking into account various
present and future telescope parameters, confusion with competing signals,
and possible systematic errors in our model. Some consequences of a future
positive detection of the signal are mentioned.


\section{Model} 
\label{sec:IGM_model} 
\indent 

Here we study the extragalactic radio signal expected from the strong intergalactic shocks involved in structure formation. We begin by reviewing the dimensional-analysis based model of Waxman \& Loeb (2000) in \S \ref{subsec:model_review}. Next, the model is generalized, and adapted for a \LCDM universe. In \S \ref{subsec:model_LCDM} we modify the halo mass function
used by the model, and incorporate the spectral features of the emitted radiation. In \S \ref{subsec:model_LCDM_sim} we calibrate the unknown dimensionless parameters of the model, using a hydrodynamical \LCDM simulation \cite{Springel2001}. The energy fractions $\xi_e$ and $\xi_B$ are evaluated in \S \ref{subsec:model_efficiency}, and their uncertainty is estimated. Finally, in \S \ref{subsec:model_halos} we compare some observational features of galaxy-cluster radio halos to the predictions of the model. 

We show that after adjusting its parameters to produce some \emph{global} features of the simulated universe, the model yields radio and \gama-ray signals consistent with those extracted independently from the simulation \cite[][in preparation] {Keshet03, Keshet04b}. The calibration scheme incorporates the inhomogeneous thermal energy injection rate along a shock, e.g. in the form of 'hot spots' at the intersections with galaxy filaments. This effectively enhances the synchrotron (but not the inverse-Compton) signal, because the magnetic energy density is higher in these regions. In addition, both synchrotron and inverse-Compton power is shifted to smaller angular scales. The integrated spectrum is shown to be nearly flat, with $\nu I_\nu$ peaking at frequencies near $100\MHz$.

In the following we use a 'concordance' $\Lambda\mbox{CDM}$ model of
Ostriker \& Steinhardt (1995) - a flat universe with normalized vacuum
energy density $\Omega_\Lambda=0.7$, matter energy density $\Omega_M=0.3$,
baryon energy density $\Omega_B=0.04$, Hubble parameter $h=0.67$, and an
initial perturbation spectrum of slope $n=1$ and normalization
$\sigma_8=0.9$. The various parameters of the cosmological model are
summarized in Table \ref{tab:CosmoParams}. 
Note that our cosmological model is slightly different from the \LCDM model used by Waxman and Loeb (2000).

\begin{deluxetable}{lll}
\tablecaption{Cosmological model and structure formation parameters. \label{tab:CosmoParams} }
\tablehead{ \em Parameter & \em Meaning & \em Value }
\startdata
$h$           & Hubble parameter & 0.67 \\ 
$k$           & Curvature & 0 \\
\tableline 
$\Omega_m$    & Matter energy density & 0.30 \\
$\Omega_{dm}$ & Dark matter energy density & 0.26 \\ 
$\Omega_b$    & Baryon energy density & 0.04 \\ 
$\Omega_\Lambda$ & Vacuum energy density & 0.70 \\
$\chi$      & Hydrogen mass fraction & 0.76 \\ 
\tableline 
$n$           & Fluctuation spectrum slope & 1 \\
$\sigma_8$    & Spectrum normalization & 0.9 \\ 
\enddata
\end{deluxetable}

\subsection{Dimensional Analysis}
\label{subsec:model_review}

Using dimensional analysis arguments, Waxman \& Loeb (2000) have related
the mass $M$ of a virialized halo, its velocity dispersion $\sigma$, its
smooth mass accretion rate $\dot{M}$ through strong shocks, its temperature
$T$ and the typical radius $r_{sh}$ of its accretion shock, by 
\beq M = \frac{\sqrt{2}}{5}\,\frac{\sigma^3(M,z)}{G H(z)} \coma \eeq 
\beqr \label{eq:model_Mdot} \dot{M}(M,z) & = & f_{acc}\frac{\sigma^3(M,z)}{G} \\ & \simeq & 2.5\times 10^{-10} f_{acc}\, h_{70}\, a^{-3/2} g(a)\, M \yr^{-1}
\coma \eeqr 
\beqr \label{eq:model_T} T(M,z) & = & f_T k_B^{-1} \,\mu \, \sigma^2
(M,z) \\ & \simeq & 1.8 \times 10^7 f_T \,h_{70}^{2/3} a^{-1} g(a)^{2/3}
M_{14}^{2/3} \K \coma \eeqr 
and 
\beqr \label{eq:model_Rsh} r_{sh}(M,z) & =
& f_{r} \frac{\sqrt{2}}{5} \, \frac{\sigma(M,z)}{H(z)} \\ & \simeq & 1.9
f_{r} \, h_{70}^{-2/3} a \, g(a)^{-2/3} M_{14}^{1/3} \Mpc \fin \eeqr 
Here $H(z) = 100 \,h \, a^{-3/2}\, g(a)\km \se^{-1} \Mpc^{-1}$ is the Hubble
parameter, with $a\equiv(1+z)^{-1}$ and $g(a)=[\Omega_m+\Omega_\Lambda a^3
+ (1-\Omega_m-\Omega_\Lambda)a]^{1/2}$, $\mu\simeq 0.65\,m_p$ is the
average mass of a particle (including electrons), $M_{14}\equiv M / 10^{14} M_\odot$, and $k_B$ is the Boltzmann constant. For consistency with previous work, we use the notation $h_{70} \equiv h/0.70$ and scale the results using $7\, \Omega_b / \Omega_m$ ($\sim 0.93$ in the present model).
The free parameters $f_{acc}$, $f_T$ and $f_{r}$ are dimensionless factors
of order unity, which are assumed (and shown in \S \ref{subsec:model_LCDM_sim}) to be roughly constant in the redshift range
and cosmological model of interest, and must be calibrated separately. The
definitions of the model parameters are summarized in Table
\ref{tab:ModelParams}, along with their values as calibrated in \S
\ref{subsec:model_LCDM_sim} and in \S \ref{subsec:model_efficiency}. 

Note that the above estimates refer to strong shocks only; weak shocks that may result from mergers of comparable mass objects are ignored since they lead to accelerated electron distributions with little energy in highly relativistic electrons \cite[for a discussion see] [] {Keshet03}. For example, merger trees have been used \cite{Gabici03} to show that the \gama-ray background from merger shocks is a factor of $> 10$ lower than the background from accretion shocks. Moreover, this emission is dominated by the strong shocks associated with mergers between halos of mass ratio $>100$, where the distinction between accretion and merger is vague \cite[see e.g.][]{Salvador98}. Weak merger events may only enhance the predicted radio background from intergalactic shocks, and only in very low photon frequencies. We shall revisit this issue in \S \ref{subsec:model_LCDM}, when discussing the radio spectrum.   

Collisionless, non-relativistic shocks are known to accelerate electrons to
highly relativistic energies. 
Keshet et al. (2003) have shown that SNR observations suggest that the energy density of relativistic electrons, accelerated by a structure formation shock, constitutes a fraction $\xi_e \simeq 5\%$ of the thermal energy density behind the shock (up to a factor of $\sim 2$, see also \S \ref{subsec:model_efficiency} for discussion). The maximal energy an electron can be accelerated to is limited by cooling, predominantly through inverse-Compton scattering of background CMB photons, yielding a maximal electron Lorenz factor $\gamma_{max} \simeq 3 \times 10^7$ \cite{LoebWaxman2000}. 

As stated above, the model focuses on strong shocks, which accelerate electrons to a power-law distribution of index $p=2$ in the differential number of accelerated electrons per electron energy (equal energy per logarithmic interval of electron energy).
The luminosity of a halo of mass $M$ at redshift $z$ due to inverse-Compton scattering of CMB photons, may thus be estimated as 
\beqr \label{eq:halo_L_IC}
\nu\,L_\nu^{iC} (M,z) & = & {1 \over {2 \ln \gamma_{max}}} \left[ {\Omega_b
\over \Omega_m} {{\dot{M} (M,z)} \over \mu} \right] \left[ \xi_e {3 \over
2} k_B T(M,z) \right] \\ & \simeq & 1.1 \times 10^{42} \left( f_{acc}\, f_T \frac{\xi_e} {0.05} \right) \left( h_{70}^{5/3} \frac{7 \, \Omega_b}{\Omega_m} \right) \nonumber \\ & & \times \left[ a^{-5/2} g(a)^{5/3} \right]
M_{14}^{5/3} \erg \se^{-1}\fin \eeqr 
Assuming that the energy density of the downstream magnetic field constitutes a fraction $\xi_B\simeq 1\%$ of the downstream thermal energy density (see \S \ref{subsec:model_efficiency}), implies that the magnetic field strength is given by 
\beqr B(M,z) & = & 0.14 \left( \frac{f_T}{f_{r}^2} \frac{\xi_B}{0.01} \right)^{1/2} \left[ h_{70}^{4/3} \left( \frac{7\,\Omega_b}{\Omega_m} \right)^{1/2} \right] \\ & & \times \left[a^{-2} g(a)^{4/3} \right] M_{14}^{1/3} \muG \coma \eeqr 
consistent with observation of galaxy cluster halos (see \S \ref{subsec:model_efficiency} for discussion). This yields a synchrotron luminosity 
\beqr \label{eq:halo_L_syn}
\nu\,L_\nu^{syn} (M,z) & = & {{B(M,z)^2/8\pi} \over u_{cmb}(z)} \,\nu\,
L_\nu^{iC} (M,z) \\ & = & 1.9 \times 10^{39} \left( \frac{f_{acc}\,
f_T^2}{f_{r}^2} \frac{\xi_e}{0.05} \frac{\xi_B}{0.01} \right) h_{70}^{13/3} \\ & & \nonumber \times \left(\frac{7\,\Omega_b}{\Omega_m}\right)^2 \left[a^{-5/2} g(a)^{13/3} \right] M_{14}^{7/3} \erg \se^{-1} \coma \eeqr
where $u_{cmb}$ is the energy density of the CMB. Note that whereas the
inverse-Compton signal depends on the parameter $\xi_B$ only
logarithmically (through the maximal energy attained by the relativistic
electrons), the synchrotron emission scales almost linearly with $\xi_B$. 

Given the number density of halos of a given mass at a given redshift, $dn(z)/dM$, one may integrate equation (\ref{eq:halo_L_IC}) or equation (\ref{eq:halo_L_syn}) to predict the inverse-Compton or the synchrotron background from intergalactic shocks,
\beq \langle \nu L_\nu \rangle = \int dz \, {{c\,dt}\over dz} \int dM {dn(z) \over dM} {{\nu L_\nu(M,z)} \over {4\pi(1+z)^4}} \fin \eeq 
For example, the Press-Schechter halo mass function gives, for our cosmological model (summarized in Table \ref{tab:CosmoParams}), an inverse-Compton background flux $\langle \nu I_\nu^{iC} \rangle = 1.3\, f_{acc}\, f_T\, (\xi_e/0.05) \ICUnits$, and a synchrotron background flux $\langle \nu I_\nu^{syn} \rangle = 3.2 \times 10^{-12} f_{acc}\, f_T^2\, f_{r}^{-2}\, (\xi_e/0.05)\, (\xi_B/0.01) \synUnits$. The halo mass function, the emitted radiation fields and the integration procedure, are all illustrated in Figure \ref{fig:PS_illustration}. 

Waxman and Loeb (2000) have calculated the two-point correlation function of the radiation emitted by intergalactic shocks. The low optical depth of large, hot clusters, which dominate the background, enables one to neglect cases where more than one cluster lies along the the line of sight, and approximate 
\beqr \label{eq:self_corr} \delta^2 I_\nu (\psi) & \equiv & \langle I_\nu(\unit{u}) I_\nu(\unit{v}) \rangle - \langle I_\nu \rangle^2 \\ & \simeq & 
\int dz \,{{c\,dt} \over dz} \int dM {dn(z) \over dM} \,{1\over {\pi r_{sh}(M,z)^2}} \, \left[ {{\nu L_\nu(M,z)} \over {4 \pi (1+z)^4}} \right]^2 \nonumber \\ & & \times \, P_{1|2} \left[ {{\psi\,d_A(z)} \over {r_{sh}(M,z)}} \right] \coma \eeqr
where $\unit{u}$ and $\unit{v}$ are unit vectors that satisfy $\unit{u} \cdot \unit{v} = \cos\psi$. The function $P_{1|2}$ is the probability that one line of sight passes through a halo of radios $r_{sh}$, given that another line of sight passes through the same halo, 
\beqr \nonumber P_{1|2}(x) & = & {2 \over \pi} \int_0^\pi d\theta \int_0^1 dy\, y \,\Theta \left[ 1-(y+x\cos\theta)^2-(x\sin\theta)^2 \right] \\ & = & \left\{ 1-{1 \over {2\pi}} \left[ x \sqrt{4-x^2} + 2 \arcsin(x/2) \right. \right. \nonumber \\ & & \left. \left. + \, 2 \arctan(x/\sqrt{4-x^2}) \right] \right\} \Theta(2-x) \coma \eeqr
where $\Theta$ is the heaviside step function. Halo-halo correlations, neglected in equation (\ref{eq:self_corr}), can only enhance the two-point correlation function. The \emph{fractional} correlation function, $\xi_\nu(\psi)\equiv \sqrt{\delta^2 I_\nu(\psi)} / \langle I_\nu \rangle$, is independent on the dynamical parameters $f_T$, $f_{acc}$, $\xi_e$ and $\xi_B$, and depends on the geometrical factor $f_{r}$ only through the relation\footnote{At low ($\nu<\nu_{br}$) or high ($\nu>\nu_{max}$) frequencies (see \S \ref{subsec:model_LCDM}), the scaling relation is more complicated.} $\xi_\nu(\psi; f_{r}) = f_{r}^{-1} \xi_\nu (f_{r}^{-1} \psi; 1)$. The contribution of halos of different mass and redshift to $\delta^2 I_\nu$ is illustrated in Figure \ref{fig:PS_illustration}.

\subsection{Adaptation for \LCDM, Spectral features}
\label{subsec:model_LCDM}

\begin{figure*}[hp]
\epsscale{1.1}
\plottwo{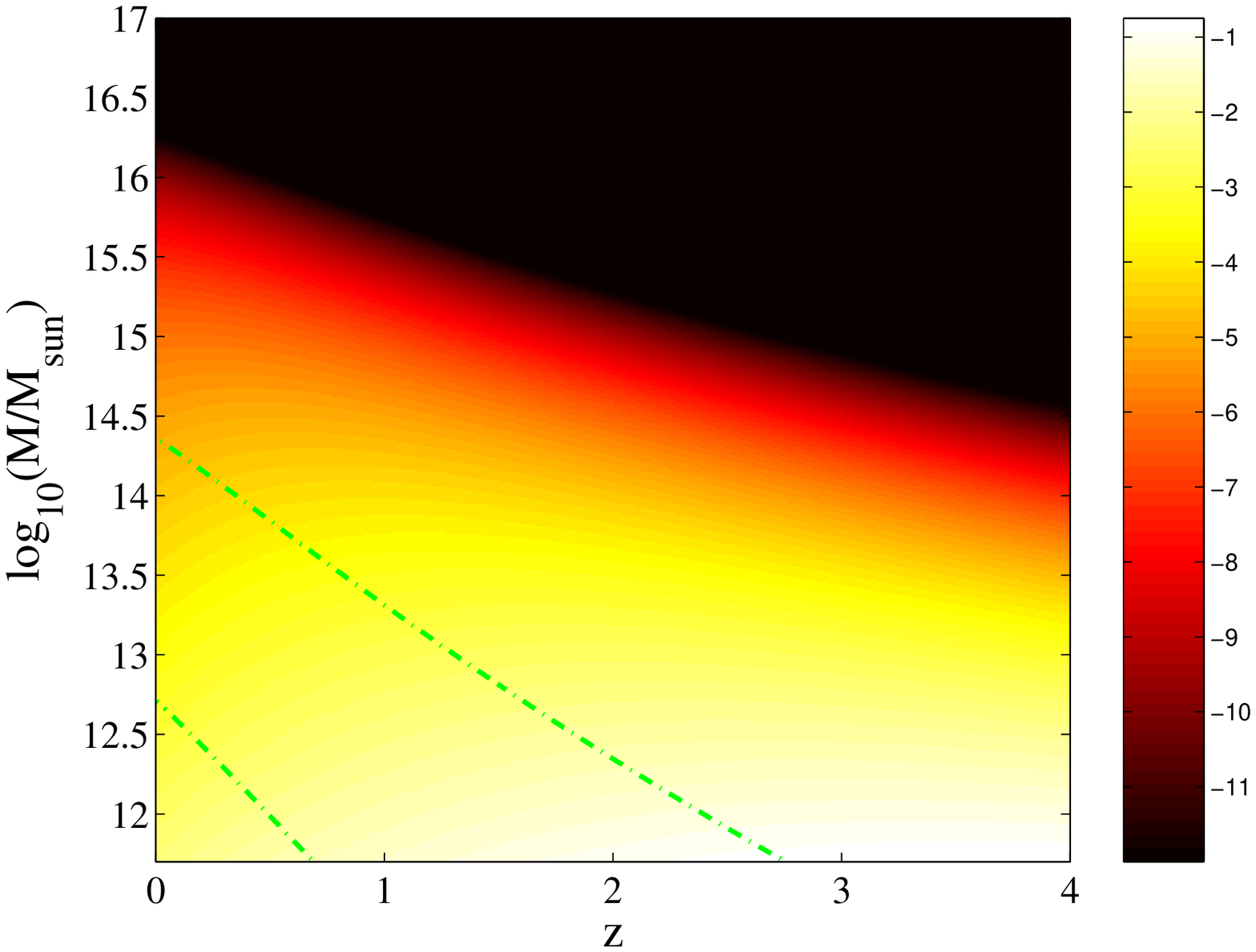}{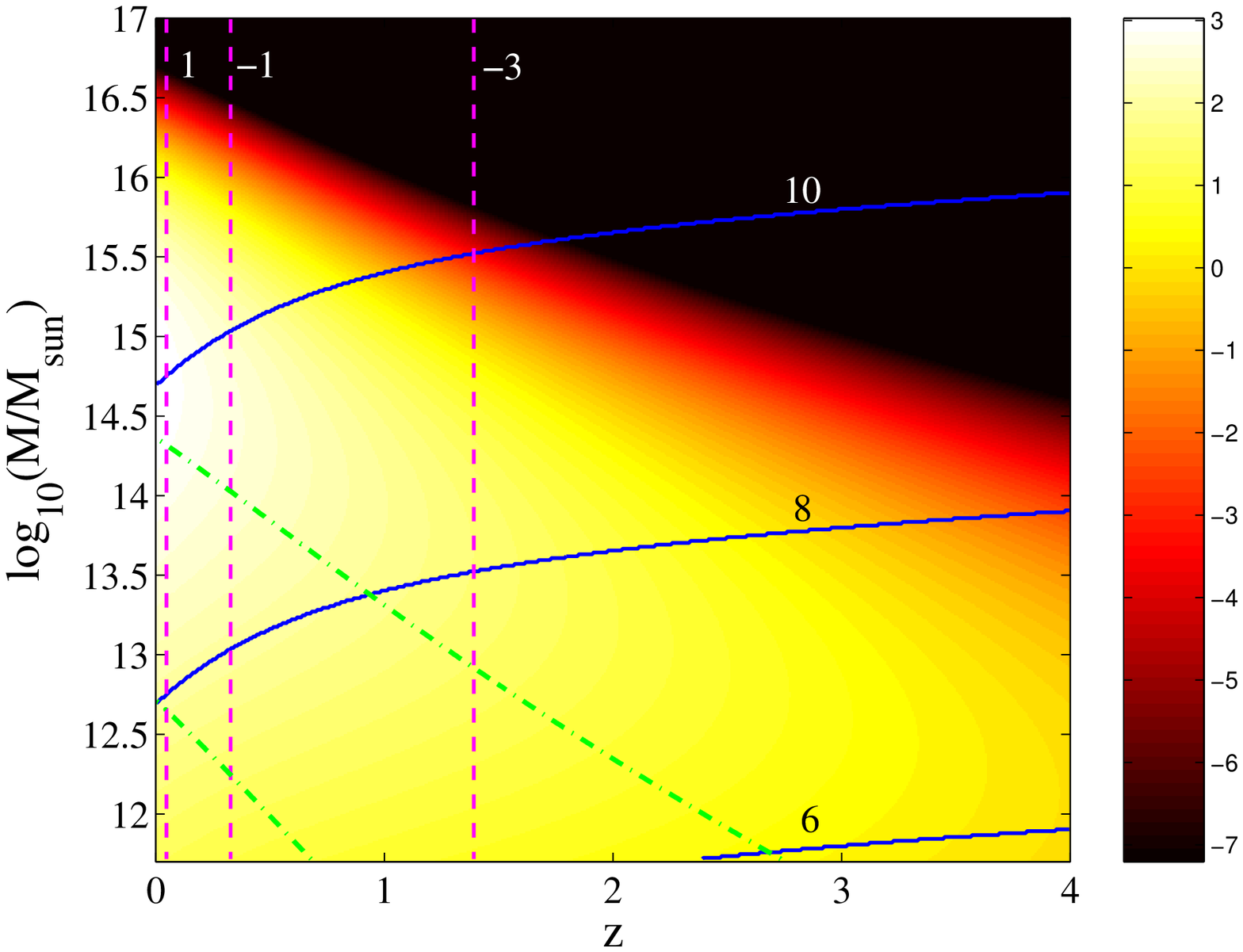}
\plottwo{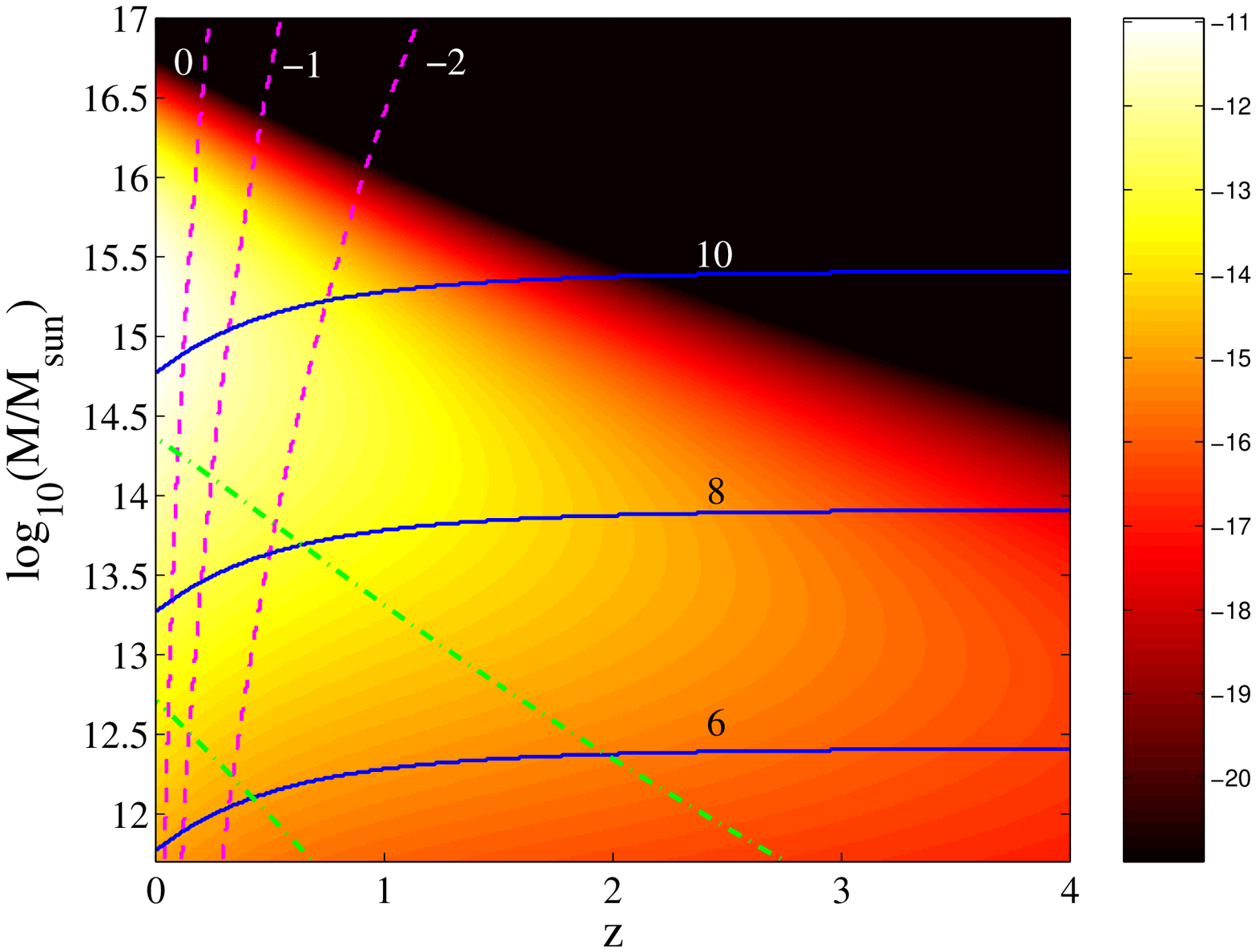}{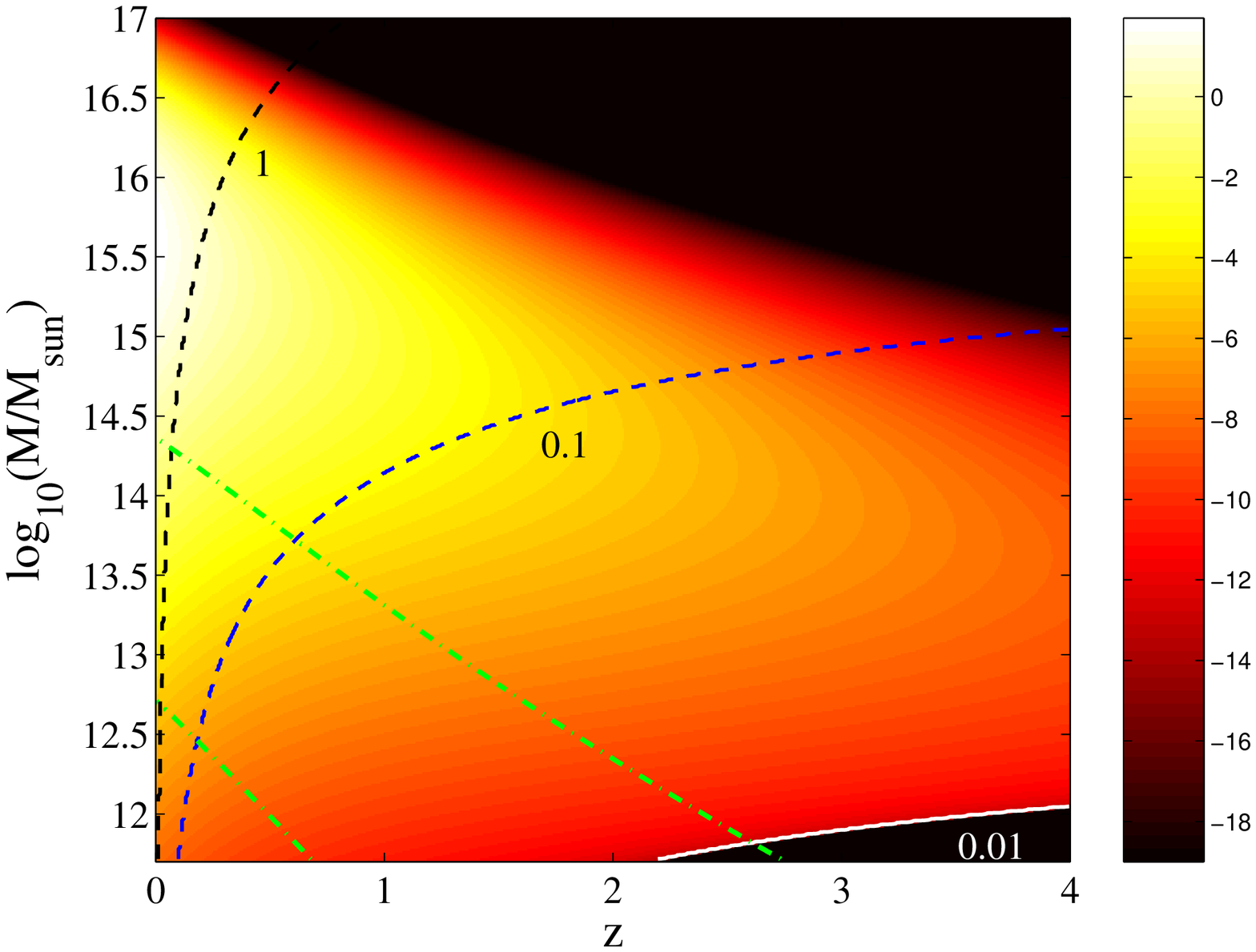}
\caption{ Illustrating the model of \S\ref{sec:IGM_model} in the redshift---halo mass plane. \newline 
Shown are the halo number density (upper-left panel, color scale: $\log_{10} \left\{ dn(z,M) / d\ln(M) \, \, [\mbox{Mpc}^{-3}] \right\}$) and the fractional halo contributions to three signals: inverse-Compton emission $\nu I_\nu^{iC}$ (upper-right, $\log_{10} f_{iC} \,[\mbox{eV s}^{-1} \cm^{-2} \sr^{-1}]$), synchrotron emission $\nu I_\nu^{syn}$ (bottom-left, $\log_{10} f_{syn} \,[\mbox{erg s}^{-1} \cm^{-2} \sr^{-1}]$), and synchrotron squared fractional two-point correlation function $\xi_\nu(\psi)^2$ (bottom-right, $\log_{10} f_{\xi^2}$). Each of the three signals is obtained from the corresponding fractional halo contribution $f$ by integration, according to $F = \int dz \int d\ln(M)\,f$. The contribution of halos in a given part of the $M$-$z$ phase space may thus be evaluated, whereas integration over the entire depicted range of $M$ and $z$ will produce results close to the total signal predicted. \newline
The results shown correspond to the modified halo mass function \cite [] [see \S \ref{subsec:model_LCDM}] {Sheth2001,Jenkins01,Barkana01}, with non-calibrated free parameters $f_T = f_{acc} = f_{r} = 1$, and energy fractions $\xi_e=0.05$ and $\xi_B=0.01$. The dash-dotted curves delineate the relatively small regions, for which fine tuning the halo mass function has \emph{lowered} the halo number density. 
Spectral features are shown for some photon frequencies for inverse-Compton (contour labels in $\log_{10} \epsilon / \mbox{keV}$) and for synchrotron emission (labels in $\log_{10}\nu/\mbox{MHz}$), as solid lines where $\nu$ coincides with $\nu_{max}(M,z)$, and as dashed lines where $\nu$ coincides with $\nu_{br}(M,z)$. 
The two-point correlation function is shown for an angular separation of $\psi=0.01\de$; for different $\psi$ values, the image is distorted and the angular cutoff (bright solid line) shifts (dashed lines, contour labels: $\psi / \mbox{deg}$).
}
\label{fig:PS_illustration}
\end{figure*}

\begin{figure}[b]
\epsscale{1.2}
\plotone{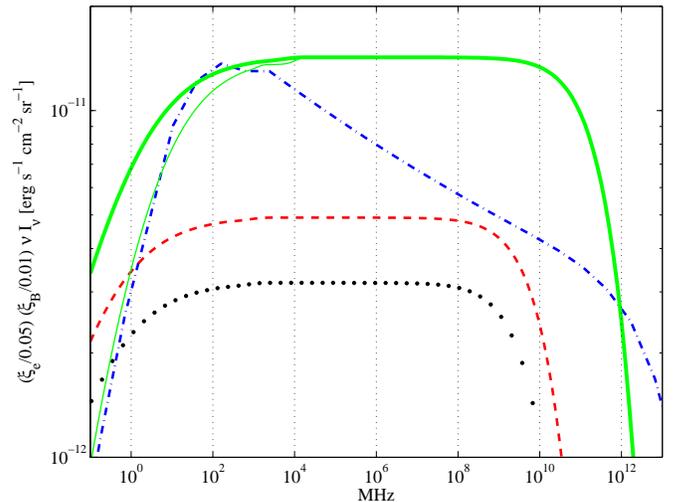}
\caption{ Spectrum of the synchrotron background from intergalactic
shocks. When using the $\Lambda$CDM-modified halo mass function, the (non-calibrated) model predicts a signal (dashed line) with $\sim 60\%$ more energy flux than found when using the Press-Schechter mass function (dotted line). The synchrotron brightness scales as $I_\nu \propto f_{acc}f_T^2 \widetilde{f}_{r}^{-2}$, so the signal corresponding to the calibrated free-parameters of Table \ref{tab:ModelParams} (heavy solid line) is a factor of $\sim 3$ stronger than the signal obtained with non-calibrated parameters, $f_T = f_{acc} = \widetilde{f}_{r} = 1$. The spectrum calculated from a \LCDM simulation \cite[dash dotted line,] [in preparation] {Keshet04b} considers only photons above the spectral break $\nu_{br}(M,z)$ of each halo. Although the simulated spectrum contains $\sim 60\%$ the energy flux of the corresponding spectrum of the calibrated model (when neglecting emission below the spectral break, thin solid line), the two spectra are similar at the relevant ($\la 10\GHz$) frequencies. 
}
\label{fig:calculated_spectrum}
\end{figure}

The Press-Schechter halo mass function is known to disagree with numerical simulations, predicting less rare, massive halos and more abundant, low mass halos. 
Sheth, Mo, \& Tormen (2001) have shown that the Press-Schechter approach agrees better with cosmological simulations, if the naive critical over-density $\delta_c(z)$ indicating collapse (and calculated for spherical collapse) is replaced by 
\beq \widetilde{\delta_c}(z,M) = \sqrt{a}\,\delta_c(z) \left\{ 1 + b \left[ {\sigma^2(M) \over {a\, \delta_c(z)^2}} \right]^c \right\} \coma \eeq
where $\sigma(M)$ is the variance of the density field smoothed on a mass scale $M$, and $a$, $b$ and $c$ are dimensionless parameters of order unity. Sheth et al. have shown that a modification of this form is naturally obtained if one considers ellipsoidal collapse and the effect of shear, although this requires that $a=1$ and does not explain the deficit in massive halos, suggesting the importance of additional effects. Numerically, good agreement with \LCDM simulations is obtained if one chooses $a=0.73$, $b=0.34$, and $c=0.81$ \cite{Jenkins01,Barkana01}. Such a modification enhances the expected background from intergalactic shocks, because it increases the density of rare, massive halos, which dominate the extragalactic signal (Figure \ref{fig:PS_illustration} illustrates the difference between the mass functions). We thus find a $20\%$ increase in the predicted inverse Compton background, $\langle \nu I_\nu^{iC} \rangle = 1.6 \, f_{acc}\, f_T\, (\xi_e/0.05) \ICUnits$. The synchrotron signal is even more sensitive to the abundance of massive halos, and is enhanced by the above modification by $50\%$, $\langle \nu I_\nu^{syn} \rangle = 4.9 \times 10^{-12} f_{acc}\, f_T^2\, f_{r}^{-2}\, (\xi_e/0.05)\, (\xi_B/0.01) \synUnits$  (see Figures \ref{fig:PS_illustration} and \ref{fig:calculated_spectrum}). In what follows, we use the modified halo number density, unless otherwise stated.

The spectrum of the integrated radiation from a halo accretion shock is essentially a broken power-law, where the spectral break is introduced by (inverse-Compton dominated) cooling, and the break frequency $\nu_{br}(M,z)$ is related to the age (or equivalently, the formation redshift) of the halo. We focus on strong shocks which accelerate electrons to a power-law energy distribution with an index $p=2$. Hence, at frequencies below the spectral break the spectrum scales as $\nu I_\nu \propto \nu^{1/2}$, reflecting the original distribution of electrons accelerated by the shock, whereas above the spectral break, cooling results in a flat ($\nu I_\nu \propto \nu^0$) spectrum. 
The spectral break frequency roughly corresponds to the minimal energy, at which electrons manage to significantly cool (say, by $\eta=50\%$ of their initial energy) between the cosmic time of their host halo accretion shock (given by its redshift $z$) and the present epoch \cite{Keshet03}, 
\beq \label{eq:nu_min} \nu_{br}(M,z) \simeq 9 \, h_{70}^2 \left( {\eta \over 1-\eta} \right)^2 {B(M,z) \over {0.1\mu\mbox{G}}} \left[ \int_0^z \frac{(1+z)^{3/2}}{g(a)}\,dz \right]^{-2} \kHz \fin \eeq 
The flat, high frequency part of the spectrum extends up to frequencies corresponding to the maximal energy of relativistic electrons, at which the latter are effectively accelerated by the shocks. The maximum Lorenz factor to which electrons are accelerated, $\gamma_{max}\simeq 3.3\times 10^7 (B/0.1\muG)^{1/2}(T/10^7\K)^{1/2}(1+z)^{-2}$ \cite{Keshet03}, thus implies maximal synchrotron frequencies around 
\beq \label{eq:nu_max} \nu_{max}(M,z) \simeq 3.6 \times 10^{14} \, \left[ {B(M,z) \over {0.1\muG}} \right]^2 {T(M,z) \over {10^7\K}} \,(1+z)^{-4} \Hz \fin \eeq 

The spectral features of the emission from intergalactic shocks may be shown pictorially in the halo mass-redshift plane (see Figure \ref{fig:PS_illustration}). As seen from equation (\ref{eq:nu_max}), low mass (cool) halos do not contribute to the spectrum at high photon frequencies, because they do not accelerate electrons to sufficiently high energies. Hence, for a given photon frequency $\nu$ we may draw a curve in the $M$-$z$ plane (the solid curves in Figure \ref{fig:PS_illustration}), such that for halos located along the curve the cutoff frequency $\nu_{max}(M,z)$ coincides with $\nu$. Halos with masses lower than found on this curve do not contribute to the integrated spectrum at $\nu$. In a similar fashion, a given photon frequency $\nu$ defines a curve in the $M$-$z$ plane (the dashed curves in Figure \ref{fig:PS_illustration}), for which the spectral break frequency $\nu_{br}(M,z)$ coincides with $\nu$. The emission from older halos (with redshift higher than found on this curve) is flat at $\epsilon$. The resulting synchrotron spectrum, integrated over all halos, is presented in Figure \ref{fig:calculated_spectrum}, and cuts off below $\sim 100\MHz$. Weak shocks, neglected in the model, accelerate softer electron distributions than discussed above, so modelling them results in an integrated spectrum slightly softer than depicted in the figure \cite[similar to the estimated inverse-Compton spectrum in the \gama-ray band, e.g.] [] {Keshet03,Miniati02,Gabici03}. Figure \ref{fig:calculated_spectrum} thus suggests, that for the total emission from intergalactic shocks (when accounting also for weak shocks), $\nu I_\nu$ peaks at frequencies around $100 \MHz$.

\subsection{Calibration Using a Cosmological Simulation}
\label{subsec:model_LCDM_sim}

We now turn to a hydrodynamical \LCDM simulation \cite{Springel2001,Keshet03} in order to obtain rough estimates of the dimensionless free parameters used in the above model. We focus on the epoch $0<z<2$, which is most relevant for radiation from intergalactic shocks. The halo parameters $f_T$, $f_{acc}$, and $f_{r}$ may be estimated by comparing various quantities calculated in the simulation, to their values according to the model. For this purpose we use global features of the simulated universe, and not the radiation fields resulting from the intergalactic shocks extracted from the simulation. This enables us to later test the calibration scheme by comparing features of the radiation from intergalactic shocks, as extracted from the simulation and as calculated from the model. Whereas the halo parameters may be calibrated in this fashion, the energy fractions $\xi_e$ and $\xi_B$ can not be similarly evaluated from a cosmological simulation, but require independent observational data (see \S \ref{subsec:model_efficiency}). 
When evaluating features of the model, we have neglected low mass halos with temperatures below $T_{min}=10^4\K$, where collapse is strongly suppressed. Our calibration scheme is insensitive to the exact value of $T_{min}$. 

The temperature parameter $f_T$ may be estimated using the temperature statistics of the baryonic component of the universe. We examine the mass averaged temperature, well fit in the simulation by $\langle T(z) \rangle_M \simeq 4\times 10^6 \,e^{-0.9z}\K$ for redshifts $0<z<2$. At low redshifts, $z<1$, the agreement between the simulation and the model is best if $f_T\simeq 0.5$. However, the redshift dependence of $\langle T(z) \rangle_M$ differs between the model and the simulation. For example, agreement between the two on the average temperature of the present-day universe, $\langle T(z=0) \rangle_M$, requires $f_T\simeq 0.45$; the best fit at the epoch $0<z<2$ is for $f_T=0.52$; and at the extreme, agreement on $\langle T(z=2) \rangle_M$ requires $f_T\simeq 0.68$. These examples imply that $f_T$ has a weak redshift dependence, and that we should consider the range of $f_T\simeq 0.45-0.55$.

In order to assess the accretion rate parameter $f_{acc}$, we examine the
fraction of mass that has been processed by {\it strong} shocks between
redshift $z_0$ and the present epoch, $f_{proc}(z<z_0)$. According to the
cosmological simulation, $f_{proc}(z<1) \simeq 18\%$, reproduced by the
model if $f_{acc}\simeq 0.12$. Here too, the redshift dependence of the
accretion rate differs between the simulation and the model, implying a
weak redshift dependence of $f_{acc}$. For example, $f_{proc}(z<2)\simeq
41\%$ corresponds to $f_{acc}\simeq 0.17$, whereas the accretion rate at very
low redshifts implies that $f_{acc}\simeq 0.08$. These findings suggest
that $f_{acc}$ lies in the range of $0.08-0.17$ for the relevant epoch.

Using the halo mass function modified for a \LCDM universe (see \S \ref{subsec:model_LCDM}) and the best fit values found for the above calibrated parameters, the model yields an inverse-Compton background $\nu I_\nu^{iC} \simeq 0.1 (\xi_e/0.05) \ICUnits$, in good agreement with the results of the simulation. Note that the result is independent of the parameters $f_r$ and $\xi_B$. Agreement to better than $\sim 30\%$ should be regarded as somewhat of a numerical coincidence, considering the uncertainties and the fact that $\nu I_\nu^{iC}$ scales linearly with the product $f_T f_{acc}$. Nonetheless, the good fit suggests that the calibration of the parameters $f_T$ and $f_{acc}$ is sensible.

Finally, the shock radius parameter $f_{r}$ can be deduced from the simulation by studying the morphology of cluster accretion shocks. A shock ring of diameter $5-10\Mpc$ has been identified around a simulated cluster of mass $M(r<5\Mpc)\simeq10^{15} M_\odot$ \cite{Keshet03}. Such a range of shock radii is obtained by the model, if $f_{r}\simeq 0.6-1.2$. Clearly, more work is required in order to obtain a better understanding of simulated cluster accretion shocks, to better calibrate $f_{r}$, and possibly identify its redshift dependence. 

An important aspect of the model is the sensitivity of the predicted synchrotron luminosity of a halo, but not its inverse-Compton luminosity, to halo asymmetry (as well as to $\xi_B$, as mentioned in \S \ref{subsec:model_review}). Such asymmetries introduce spatial fluctuations in the thermal energy injection rate through the halo accretion shock. Whereas the inverse-Compton emission from a particular region along the shock front scales linearly with the downstream thermal energy, the synchrotron emission scales as the \emph{square} of the thermal energy, $\nu L_\nu^{syn} \propto \dot{M} T^2$. This implies, that fluctuations along the shock front will increase the overall synchrotron luminosity of the halo, leaving its inverse-Compton luminosity intact. Accounting for fluctuations is crucial, because cosmological simulations \cite[e.g.][]{Minitai01} find highly asymmetric shocks, with very strong fluctuations along the shock fronts. In particular, the simulation we have studied \cite{Keshet03} reveals "bright spots" in the thermal energy injection rate at the intersections of the accretion shock of a $\sim 10^{15} M_\odot$ cluster with large galaxy filaments, channelling large amounts of gas into the cluster region. The thermal injection rate (and the resulting \gama-ray brightness) of these regions is more than an order of magnitude higher than the typical brightness along the shock front \cite[see][Figures 9 and 10]{Keshet03}. 

The large fluctuations in thermal injection rate necessitate the
introduction of a geometrical correction factor into the simple model
presented in \S \ref{subsec:model_review}. We note the localized nature of
the hot regions along the shock front, and the large ratio $\zeta\simeq 10$
between the temperature of these regions and the temperature of the dimmer,
more extended regions. These features justify approximating the geometrical
correction by attributing all the emission from a halo accretion shock to
an effective smaller region, of scale $\widetilde{r}_{sh} \simeq
r_{sh}/\zeta$. This implies replacing the dimensionless parameter $f_{r}$
with a different parameter, $\widetilde{f}_{r} \equiv f_{r}/ \zeta$. Note
that such a replacement will enhance the synchrotron luminosity of the halo
by a factor $\zeta^2$, while leaving its inverse-Compton luminosity
unchanged. As an order of magnitude estimate, $f_{r}\simeq 0.6-1.2$
roughly yields $\widetilde{f}_{r}=f_{r}/\zeta \simeq 0.05-0.20$, with a
middle-of-the-road estimate of $\widetilde{f}_{r} = 0.1$. Note that the
geometrical correction alters the two-point correlation function of
both synchrotron and inverse-Compton emission, shifting power to smaller
angular scales.

\begin{figure}[h]
\epsscale{1.2}
\plotone{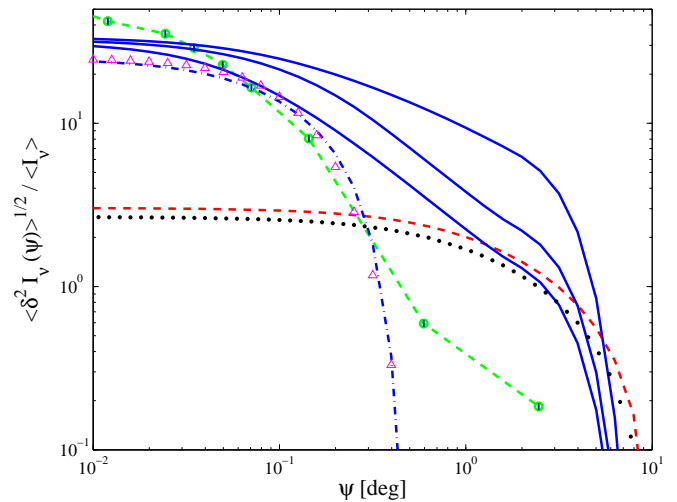}
\caption{ Fractional two-point correlation function $\xi_\nu(\psi,\nu)$ of synchrotron background from intergalactic shocks, for $\nu=100\MHz$. With the modified halo mass function, the (non-calibrated) model predicts a signal (dashed line) slightly stronger than when using the Press-Schechter mass function (dotted line). The result of the calibrated model (dash-dotted line), scaling according to $\xi_\nu(\psi,\nu; \widetilde{f}_{r}) = \widetilde{f}_{r}^{-1} \xi_\nu (\widetilde{f}_{r}^{-1} \psi,\nu; 1)$ for $\nu_{br} \ll \nu \ll \nu_{max}$, has more power on small scales. These model predictions are well fit by the functional form $\xi_\nu(\psi, \nu) = A(\nu) \exp \left\{ -[\psi/\psi_0(\nu)]^{q(\nu)} \right\}$, e.g. a fit is shown (triangles) for the calibrated model signal, with $A=25$, $\psi_0=9^\prime$ and $q=1.5$. The \LCDM simulation signal \cite[circles with dashed line to guide the eye, error bars represent statistical errors introduced by averaging over random pairs of lines of sight, see] [] {Keshet04b} is similar to the calibrated model result, and yields even more small scale power. The above results do not include emission below the break frequency $\nu_{br}$. The results of the calibrated model for the entire spectrum have more large-scale power, due to the contribution of young, nearby halos, and are presented for frequencies $1\MHz$, $100\MHz$ and $10\GHz$ (solid lines, bottom to top).  
}
\label{fig:calculated_correlation}
\end{figure}

The model parameters calibrated above are summarized in Table \ref{tab:ModelParams}. We present the resulting synchrotron radio background in Figure \ref{fig:calculated_spectrum}, and the synchrotron two-point correlation function in Figure \ref{fig:calculated_correlation}. It is encouraging to note that after calibrating the free parameters with essentially global features of the cosmological simulation, the model predictions agree well with results extracted from the simulation, regarding the inverse-Compton \gama-ray background \cite{Keshet03}, the synchrotron radio background, and the synchrotron two-point correlation function \cite{Keshet04b}. A more accurate calibration of the parameters, including an evaluation of their redshift dependence, may be obtained from a detailed analysis of the various clusters identified in cosmological simulations.

\begin{deluxetable}{lll}
\tablecaption{\label{tab:ModelParams}
Parameters of the intergalactic shock model.}
\tablewidth{0pt}
\tablehead{ \em Param. & \em Quantity parameterized & \em Value (Range)}
\startdata 
$f_T$     & Halo downstream temperature \tablenotemark{a} & 0.50 (0.45-0.55) \\ 
$f_{acc}$ & Halo mass accretion rate \tablenotemark{a}   & 0.12 (0.08-0.17) \\ 
$f_{r}$  & Halo shock radius \tablenotemark{a}          & 0.9 (0.6-1.2) \\ 
$\widetilde{f}_{r}$ & Halo energy injection scale \tablenotemark{a} & 0.1 (0.05-0.2) \\
\tableline
$\xi_e$ & Energy fraction in relativistic electrons \tablenotemark{b} & 0.05 (0.02-0.10) \\
$\xi_B$ & Energy fraction in magnetic fields \tablenotemark{b} & 0.01 (0.005-0.04) \\
\enddata
\tablenotetext{a}
{Defined in \S \ref{subsec:model_review} and calibrated in \S \ref{subsec:model_LCDM_sim} using a hydrodynamical cosmological simulation. }
\tablenotetext{b}
{Energy fractions out of the shock thermal energy, evaluated in \S \ref{subsec:model_efficiency} according to observations of SNRs and galaxy cluster halos. }
\end{deluxetable}

\subsection{Energy Conversion Efficiency}
\label{subsec:model_efficiency}

Intergalactic shock waves are expected to accelerate electrons to highly
relativistic energies, and to strongly amplify magnetic fields. The average
fractions of shock thermal energy deposited in relativistic electrons
($\xi_e$) and in magnetic fields ($\xi_B$) are both important parameters of
our model, each bearing linearly upon the predicted synchrotron signal, and
thus deserve a special discussion here.

\subsubsection{Electron acceleration}

Collisionless, non-relativistic shock waves are known to Fermi accelerate a power-law energy distribution of relativistic particles. This phenomenon has been observed in astrophysical shock waves on various scales, such as in shocks forming when the supersonic solar wind collides with planetary magnetospheres, in shocks surrounding SNRs in the interstellar medium, and probably also in shocks in many of the most active extragalactic sources, quasars and radio galaxies \cite{Drury83,Blandford87}. The electron power-law distributions extend up to $\sim 100 \TeV$ energies in SNRs \cite{Tanimori98}, where shock velocities are of order $v\sim 10^3 \km\se^{-1}$, similar to intergalactic shock velocities. 
 
Although no existing model credibly calculates the acceleration efficiency,
a simple argument \cite{Keshet03} suggests that we may evaluate $\xi_e$ in strong intergalactic shocks using the estimated acceleration efficiency of other strong astrophysical shock waves. Consider an ideally strong, non-relativistic shock wave, such that the shock Mach number $\Upsilon \gg 1$ and the thermal energy of the upstream plasma is negligible with respect to the shock energy. The physics of such a shock is essentially determined by three parameters, namely the shock velocity $v$, the upstream plasma number density $n_u$, and the upstream magnetic field strength $B_u$ (in principle, the result may also depend on the detailed structure of the upstream magnetic field). The upstream density may be eliminated from the problem altogether by measuring time in units of $\nu_{p}^{-1}$, where $\nu_{p}$ is the plasma frequency. The upstream magnetic field strength, parameterized by the cyclotron frequency $\nu_{c}$, can not be additionally scaled out of the problem. However, comparing $\nu_{c,i}$ (where subscript $i$ denotes a property of the ions) to the growth rate of electromagnetic instabilities in the shocked plasma, $\nu_{ins} = \nu_{p,i} \,v /c$, indicates that their ratio in strong shocks satisfies $(\nu_{c,i}/\nu_{ins})^2 = (B_u^2/8\pi) / (n_u m_p v^2/2) \ll 1$. We thus \emph{assume} that there is a well behaved limit when this ratio approaches zero, implying that the upstream magnetic field has little effect on the characteristics of strong shocks. With this assumption, we expect to find much similarity between sufficiently strong shocks in different environments, provided that their shock velocities are comparable, regardless of the plasma density and the strength or structure of the upstream magnetic field. The little effect of upstream magnetic fields on strong shocks is supported by recent observations of SNR shocks, which suggest that the shocks produce strongly fluctuating, near equipartition (with respect to the thermal energy of the non-relativistic electrons, hereafter) magnetic fields, much stronger than the magnetic field far upstream \cite[and discussion of magnetic field amplification, below]{Bamba03,Berezhko03}.

The preceding discussion indicates that the best analogy to strong intergalactic shocks may be found in the strong SNR shocks, drawing upon the similarity between the velocities of the two families of shocks, both of order $v \simeq 10^3 \km \se^{-1}$. The fraction of shock energy deposited in relativistic electrons by SNR shocks was estimated by several authors, with the most reliable estimates found for remnants with \gama-ray detection such as SNR1006 \cite{Tanimori98}. Dyer et al. (2001) have modelled the multi-frequency emission from SNR1006, finding an acceleration efficiency of $\xi_e=5.3\%$ (corresponding to a fraction $\eta_e \simeq 1.4\%$ out of the total supernova explosion energy). Other authors \cite{Mastichiadis96, Aharonian99} have estimated $\eta_e\simeq 1\%-2\%$ in SNR1006, which corresponds to $\xi_e$ values in the range of $3.8\%-7.6\%$. Recent \emph{Chandra} observations, resolving thin sheets in the NE shock front of SNR1006, suggest electron acceleration more localized and more efficient than previously thought, with near equipartition energy (with respect to the thermal energy of the non-relativistic electrons) in relativistic electrons \cite{Bamba03}. The extrapolated value of $\xi_e$ depends on the uncertain magnetic field strength; the lowest plausible energy fraction in relativistic electrons corresponds to an equipartition magnetic field, for which $\xi_e \simeq 4\%-11\%$ for different sheet regions. Ellison, Slane and Gaensler (2001) have modelled the emission from SNR G$347.3-0.5$, estimating the energy of relativistic electrons to constitute at least $1.2\%$, and more likely $2.5\%$, of the shock kinetic energy flux, corresponding to $2.5\% \la \xi_e\simeq 5\%$.

The above estimates of the acceleration efficiency of SNR shocks, suggest
that for strong intergalactic shocks $\xi_e\simeq 0.05$, with an
uncertainty factor of $\sim 2$. An independent, less reliable method for
estimating $\xi_e$ relies on the ratio between the energies of cosmic-ray
electrons and cosmic-ray ions in the interstellar medium, suggesting that
$\xi_e \simeq 1\%-3\%$ \cite[for a discussion, see][]{Keshet03}. However,
since the relation between this ratio in the interstellar medium and
immediately behind intergalactic (or SNR) shocks is unknown, this estimate
is highly uncertain. We stress that our estimate of $\xi_e$ is based only
on \emph{observations} of SNRs, without provoking any elaborate model for
the acceleration mechanism of the electrons by the shocks. The high
efficiency of electron acceleration deduced, suggests that a substantial
fraction of the shock thermal energy is transferred into relativistic ions,
indicating that a non-linear theory \cite[e.g.] [and references therein]
{Berezhko99} is required in order to account for the shock structure and
the particle acceleration process. Non-linear theories for diffusive
acceleration of particles by shock waves are at the present stage
incomplete \cite[for a recent review, see] [] {Malkov01}. 
In particular, such models predict deviations of the accelerated particle distribution from a pure power-law. Our model, which is based on observations of SNR shocks, does not incorporate such deviations.

\subsubsection{Magnetic field amplification}

Magnetic fields in galaxy cluster halos have been estimated using several
techniques, based on diffuse synchrotron emission from the cluster,
preferably combined with inverse-Compton detection, based on Faraday
rotation of background or embedded polarized radio sources, and based on
the observation of cold fronts in cluster X-ray images \cite[for reviews,
see] [] {Kronberg94, Henriksen98, Carilli02}. The synchrotron emission from
a cluster measures the volume-averaged magnetic field weighted by the
relativistic electron distribution, and is thus most relevant for our
model. Studies of the resulting radio signal suggest volume averaged
magnetic field strengths of the order of a few $0.1\muG$, close to the
values inferred for a minimal energy configuration with equal energy in
magnetic fields and in relativistic electrons. Faraday rotation and cold
front studies suggest stronger magnetic fields of several $\mu$G, but are
sensitive to different measures of the magnetic field. Faraday rotation
studies, for example, suggest magnetic fields of several $\mu$G with a high
filling factor up to $\sim 0.5 \Mpc$ from the cluster center
\cite{Clarke01}. However, such studies are sensitive to the magnetic field
weighted by the \emph{thermal} electron distribution, suffer from an
uncertain level of depolarization internal to the source, and their
analysis depends on the assumed configuration of the magnetic field.

In the Coma super-cluster, as the most studied example, magnetic field measurements based on the cluster-size radio halo \cite{Kim89}, excess emission in the X-ray \cite{Rephaeli99,Fusco-Femiano99} and in the extreme UV \cite[][but see also Bowyer et al. 1999]{Hwang97} bands, and lack of detection in the \gama-ray band \cite{Sreekumar96}, all suggest a volume-averaged magnetic field of a few $0.1\muG$, close to the value of $\sim 0.4\muG$ obtained by assuming equal energy in magnetic fields and in relativistic electrons \cite{Giovannini93}. Studies of Faraday rotation of background polarized sources, on the other hand, suggest magnetic fields of several $\muG$, entangled on $<1\kpc$ scales \cite{Kim90,Feretti95}. 

A fraction $\xi_B\simeq0.01$ of a shock thermal energy transferred to magnetic energy, implies magnetic field strengths that are an order of magnitude lower than their equipartition value, and are close to their value in a configuration with equal energy in magnetic fields and in relativistic electrons (assuming $\xi_e \simeq 0.05$). In cluster halos, our calibrated model reproduces, for this choice of $\xi_B$, a volume-averaged magnetic field strength $B\simeq 0.1\muG$ (for $M\simeq 10^{14}M_\odot$), although the 'bright spots' could contain magnetic field strengths as high as $1\muG$. The lowest observational estimates for magnetic field strengths in such clusters are $\ga 0.05\muG$, suggesting that $\xi_B$ could be higher than our chosen value by a factor of a few, but is unlikely to be smaller than it by more than a factor of $\sim 2$.

It is important to note that recent studies suggest that SNRs contain
strong, near equipartition magnetic fields. For example, Pannuti et
al. (2003) claim that modelling the broad band (radio to \gama-ray)
observations of SNR G347.3-0.5 requires a magnetic field strength $B
\simeq 150_{-80}^{+250} \muG$, and implies highly localized X-ray
emission. The existence of strong, near equipartition magnetic fields
behind SNR shocks has recently been confirmed by the highly localized
nature of hard X-ray emission in the resolved sheets composing the shock front
of SNR1006 \cite[e.g.][]{Berezhko03}. The narrow extent of emission observed upstream of these sheets implies a small upstream diffusion constant, corresponding to upstream magnetic fields of strength $\ga 10\muG$, with strong fluctuations ($\delta  B/B\simeq 1$) on small, $d \ll 10^{17}\cm$ scales \cite{Bamba03}. Such magnetic field fluctuations are far stronger, and have much smaller scales, than found in the surrounding ISM, and so must be induced by the shock. These conclusions strongly suggest that the assumption, that the physics of a strong shock depends only weakly upon the far upstream magnetic field, is valid, justifying the analogy between strong intergalactic shocks and strong SNR shocks of comparable velocities.

\subsection{Comparison with Radio Halo Observations}
\label{subsec:model_halos}

\begin{figure}[h]
\epsscale{1.2}
\plotone{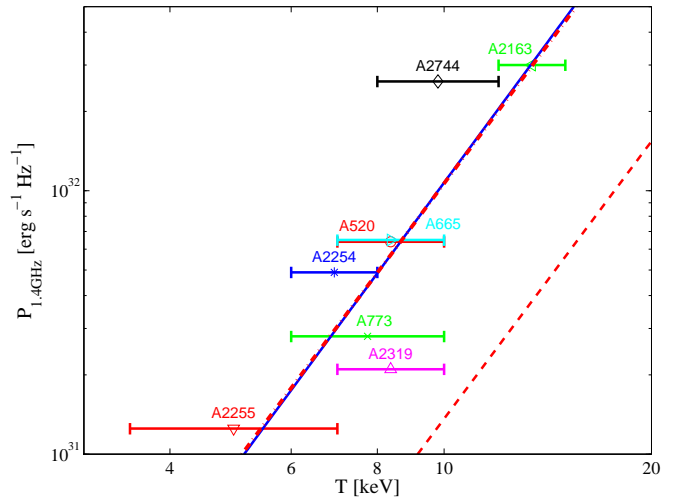}
\caption{ Specific luminosities of galaxy-cluster radio halos for $\nu = 1.4\GHz$, plotted against the temperature ranges of their host clusters \cite[error bars with cluster names, adopted from] [] {Govoni01, Feretti03}. The best power-law fit to the data (solid line) scales as $L_\nu (T,\nu = 1.4 \GHz) \propto T^{3.55}$. The specific luminosity according to the model scales similarly, $L_{\nu}(T) \propto T^{7/2}$, but the normalization (of the calibrated model, dashed line) is different; the dash-dotted line presents the result of the model where the halo temperature was re-scaled by a factor of $1.8$. 
}
\label{fig:clusters}
\end{figure}

Radio halos are observed in $\sim 35\%$ of the young, massive galaxy clusters \cite[with X-ray luminosity $L_X>10^{45}\erg\se^{-1}$, see] [] {Giovannini99, Feretti03}, and their radio luminosity is known to be correlated with the cluster temperature \cite{Liang01} and mass \cite{Govoni01}. Figure \ref{fig:clusters} shows the specific luminosities of such halos for $\nu = 1.4 \GHz$, plotted against the temperature ranges of their host clusters. A simple power-law fit to the data, $L_\nu(T, \nu=1.4\GHz) \propto T^{\phi}$, gives $\phi=3.55$, which is very similar to the power-law index $\phi=7/2$ predicted by our model, although the (calibrated) model yields a specific luminosity that is lower than observed by a factor of $\sim 8$ (equivalently, the temperatures are discrepant by a factor of $\sim 1.8$). A similarly good qualitative agreement between the observations and the model is obtained when replacing the cluster temperatures with their estimated masses, giving $L_\nu (M, \nu=1.4 \GHz) \propto M^{2.2}$ \cite{Govoni01}, similar to the model prediction $L_\nu^{syn} (M) \propto M^{7/3}$. However, the estimated mass of a cluster is somewhat less certain than the measured temperature, and is sensitive to the definition of the cluster boundary.

Some comments regarding Figure \ref{fig:clusters} are in place. First note, that the agreement between model and observations regarding the value of $\phi$ is insensitive to the free parameters of the model. The specific synchrotron luminosity of the model, when written as a function of halo temperature, scales linearly with the combination $f_{acc} f_{r}^{-2} f_T^{-3/2} \xi_e \xi_B$. The strong dependence of this result upon the model parameters, and the low redshifts ($z \la 0.3$) of all the halos depicted in Figure \ref{fig:clusters}, suggest that the discrepancy between model and observations may be eliminated by better calibrating the parameters, and modelling their redshift dependence (note that this will \emph{increase} the radio signal we predict). Second, note that the low fraction of clusters with observed radio halos could result from a combination of two effects: (i) not all clusters have been significantly accreting new mass at the cosmic time when they are observed; and (ii) the surface brightness of the halos observed is low, close to the instrumental sensitivity, suggesting that more radio halos will be observed with future telescopes. A varying accretion rate will tend to enhance the luminosity of the halos that are observed, possibly accounting for some of the discrepancy seen in Figure \ref{fig:clusters}. Finally, note that Miniati et al. (2001) have found, using a cosmological simulation, that $\phi=2.6-2.8$ (for emission from electrons accelerated at shocks; for emission from secondary electrons produced by $p$-$p$ collisions of cosmic ray ions, they found $\phi=4.1-4.2$). However, their simulation differs substantially from our model and from the simulation of Keshet et al. (2003, 2004b), regarding their cosmological model (SCDM), electron acceleration efficiency \cite[$\xi_e< 0.5\%$, see also] [] {Miniati02}, and magnetic field normalization ($\langle B^2 \rangle^{1/2} \simeq 3\muG$ for a Coma-like cluster).


\section{Observational Consequences} 
\label{sec:feasibility}
\indent

In this section we examine the capability of present-day and next generation radio telescopes operated from the ground (the LOFAR and the SKA) and from space (the ALFA) to detect the radio emission from intergalactic shocks. The LOFAR is an interferometric imaging telescope for the $10-240 \MHz$ band, planned to begin initial operation on $2006$. The SKA is a square kilometer interferometric array for the $150\MHz-20\GHz$ frequency range, planned to become operational during the next decade. The ALFA mission is a proposed space interferometer composed of 16 satellites, for very low frequencies in the range $30\kHz-30\MHz$. The main features of these telescopes are summarized in Table \ref{tab:TelescopeParams}. 

This section is organized as follows. In \S
\ref{subsec:feasibility_notations} we present the notations used throughout
the section. Various foreground and background signals are discussed in \S
\ref{subsec:feasibility_signals}, in an attempt to assess the optimal
conditions for detecting the emission from intergalactic shocks, with as
little contamination as possible. We summarize the results of this section
in \S \ref{subsec:feasibility_summary}, and discuss scenarios by which the
signal could be identified. A detailed analysis of low-frequency sky-brightness observations and their interpretation will be given in \S
\ref{sec:LFRB}. Implications of the model uncertainties on the feasibility
of detecting the signal, will be discussed in \S\ref{sec:discussion}.

\begin{deluxetable*}{cccc}
\tablecaption{\label{tab:TelescopeParams}
Parameters of next generation radio telescopes.}
\tablewidth{0pt}
\tablehead{ \em Parameter & \em LOFAR & \em SKA & \em ALFA }
\startdata 
Frequency range & 10 MHz - 90 MHz & 150 MHz - 20 GHz & 30 kHz - 30 MHz \\
  & 110 MHz - 240 MHz &              &               \\
\tableline  
Angular resolution & $<10\arcsec$ near 15 MHz & $<0.1\arcsec$ for $1.4\GHz$ 
                & $1\de.7$ near 100 kHz  \\
                & $<1\arcsec$ near 150 MHz &                           
                & $1\arcmin$ for 10 MHz  \\
\tableline  
Continuum surface & $10\mK$ on $5\arcmin-10\arcmin$ in & $1\K$ on $0.1\arcsec$ & \nodata \\
brightness sensitivity & high frequency band    &  & \\
\enddata
\end{deluxetable*}

\subsection{Notations}
\label{subsec:feasibility_notations}
\indent 

As mentioned in \S \ref{sec:IGM_model}, the two-point correlation function of specific intensity fluctuations at a given angular separation $\psi$, is defined as 
\beqr \delta^2 I_\nu (\psi) & \equiv & \langle \delta I_\nu(\unit{u}) \delta I_\nu(\unit{v}) \rangle \\ & = & \langle I_\nu(\unit{u}) I_\nu(\unit{v}) \rangle - \langle I_\nu \rangle^2 \coma \eeqr
where $\unit{u}$ and $\unit{v}$ are unit vectors that satisfy $\unit{u} \cdot \unit{v} = \cos\psi$, and $\delta I_\nu (\unit{u}) \equiv I_\nu(\unit{u}) - \langle I_\nu\rangle$. The specific intensity is related to the brightness temperature $T_b$ (which approximately equals the thermodynamic temperature for the sky brightness and the frequency range of interest) by $I_\nu = 2 \nu^2 k_B T_b /c^2$. The following discussion of intensity fluctuations is thus equally applicable for temperature fluctuations, up to a multiplicative constant. 

\begin{figure}[h]
\epsscale{1.2}
\plotone{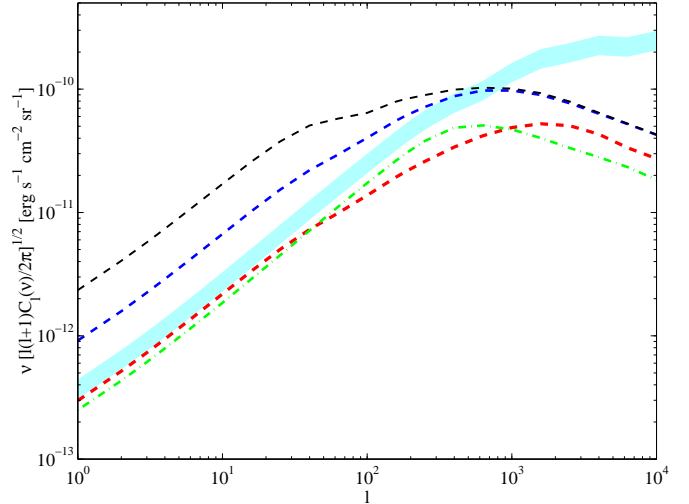}
\caption{ Angular power spectrum of the synchrotron background from intergalactic shocks. The logarithmic contribution to the variance $\nu \delta I_l= \nu [l(2l+1) C_l / 4\pi]^{1/2}$ is shown according to the calibrated model (see \S \ref{subsec:model_LCDM_sim} and Table \ref{tab:ModelParams}) for frequencies $1\MHz$, $100\MHz$, and $10\GHz$ (dashed lines, bottom to top). The result of the \LCDM simulation \cite{Keshet04b} for $\nu=100\MHz$ (band, with thickness corresponding to the statistical error), neglecting emission below the spectral break, exhibits more power than the corresponding result of the calibrated model (when neglecting emission below the spectral break, dash dotted line), in particular on small, arcminute angular scales.  }
\label{fig:synch_Cl}
\end{figure}

It is often advantageous to study the \emph{Angular Power Spectrum} (hereafter APS) of various signals, in order to obtain a direct estimate of their importance at various angular scales. The angular power spectrum $C_l$ is defined through the relation 
\beq \label{eq:corr_func} \delta^2 I_\nu (\psi) \equiv \frac{1}{4\pi} \sum_{l=1}^{\infty} (2l+1)C_l(\nu) \,P_l(\cos\psi) \coma \eeq
where $P_l(x)$ is the Legendre polynomial of degree $l$. 
The power at a given multipole $l$ is often expressed using its logarithmic contribution to the intensity variance, $\delta I_l \equiv [l(2l+1)C_l/4\pi]^{1/2}$ (or $\delta T_l$ for the variance of the brightness temperature), where multipole $l$ roughly corresponds to angular scales $\theta \simeq 180\de/l$. Using the orthogonality of the Legendre polynomials, we may invert equation (\ref{eq:corr_func}) to find the APS, 
\beq \label{eq:Cl_from_xi} C_l(\nu) = 2\pi \int_0^\pi \delta^2 I_\nu(\psi) P_l(\cos \psi) \sin \psi \,d \psi \fin \eeq
The APS of synchrotron emission from intergalactic shocks may be calculated from the two-point correlation function (see Figure \ref{fig:calculated_correlation}) using equation (\ref{eq:Cl_from_xi}). The resulting APS, presented in Figure \ref{fig:synch_Cl}, peaks at multipoles $l\simeq 400-4000$ for the relevant frequency range, corresponding to angles $\theta \simeq 3^\prime-30^\prime$.

In order to illustrate the meaning of the APS, we expand the fluctuations in the specific intensity by the spherical harmonics $Y_{lm}(\unit{\sigma})$, 
\beq \delta I_\nu(\unit{u}) = \sum_{l=1}^\infty \sum_{m=-l}^l a_{lm}(\nu) Y_{lm}(\unit{\sigma}) \fin \eeq 
A radiation field drawn from an \emph{isotropic} distribution satisfies 
\beq \langle a_{lm}(\nu) \rangle = 0 \eeq
and 
\beq \label{eq:isotropic_feature2} \langle a_{lm}^*(\nu) a_{l^\prime m^\prime} (\nu) \rangle = \delta_{ll^\prime}\delta_{mm^\prime} C_l(\nu) \coma \eeq 
where $\langle\rangle$ represents averaging over the ensemble. The Legendre polynomial addition theorem justifies the identification of $C_l$ as the APS in equation (\ref{eq:isotropic_feature2}). In practice, for signals which are not necessarily isotropic, such as the Galactic foreground, it is traditional \citep[e.g.][]{Tegmark96} to define  
\beq C_l(\nu)\equiv \frac{1}{2l+1}\sum_{m=-l}^l \langle |a_{lm}(\nu)|^2\rangle
\fin \eeq
It is often advantageous to analyze extragalactic signals by focusing on a small region $P$ in the sky, where Galactic foreground is minimal. When studying angular correlations within $P$, the contribution of multipoles that correspond to angular scales larger than $P$ is diminished. Hence, for angular scales $\theta$ much smaller than $P$, the correlation function may be approximated using equation (\ref{eq:corr_func}), by introducing a low multipole cutoff at $l(P)$. For example, the correlation function in a sky patch $P$ of typical angular size $15\de$ ($\sim0.4\%$ of the sky) involves only multipoles $l>l(P)\simeq 10$. At the other extreme, a high multipole cutoff is introduced by the angular resolution of the experimental device. Assuming a Gaussian response function with standard deviation $\theta_c$, this is incorporated into equation (\ref{eq:corr_func}) as an exponential multipole cutoff. With these modifications, we conclude that 
\beq \label{eq:corr_func2} \delta^2 I_\nu (\psi) = \frac{1}{4\pi} \sum_{l=l(P)}^{\infty} (2l+1)C_l(\nu) \,P_l(\cos\psi) \exp{ \left( - \frac{l^2\theta_c^2}{2} \right)} \fin \eeq

Finally, the sensitivity of a telescope is often estimated using its RMS noise level for a given beam size $\sigma = \pi \psi^2$, denoted $\delta I_{rms} \equiv [\delta^2 I(\sigma)]^{1/2}$. For well behaved signals, where the two-point correlation function $\delta^2 I(\psi)$ is a monotonically decreasing function of $\psi$, one finds $\delta^2 I(\psi)<\delta^2 I_{rms} (\sigma)$. Hence, the RMS noise level sensitivity imposes an upper limit to the two-point correlation function sensitivity, and may be compared directly with the logarithmic contribution to the variance at the corresponding multipole, $\delta I_l$.

\subsection{Competing Signals}
\label{subsec:feasibility_signals}
\indent

In the following, we examine various foreground and background signals in the radio band, for the relevant frequency range ($1\MHz\la \nu \la 10\GHz$). We begin with Galactic synchrotron foreground and with the integrated signal from discrete radio sources, which contaminate the expected fluctuating signal from intergalactic shocks on large and on small angular scales, respectively. We then discuss other extragalactic radio signals of interest, namely bremsstrahlung from \Lya clouds and 21 cm tomography. For each signal discussed, we present its contribution to the sky brightness in Figure \ref{fig:LFRB_sky}, and its logarithmic contribution to the variance on a $0\de.5$ angular scale in Figure \ref{fig:Sky_Cl}.

\subsubsection{Galactic Synchrotron Emission} 
\indent

At low frequencies $\nu \la 1\GHz$, the brightness of the sky is dominated
by Galactic synchrotron emission, produced by cosmic-ray electrons gyrating
in the magnetic fields of the interstellar medium (ISM). At low Galactic latitudes, there is also a substantial contribution of free-free emission from low-latitude \HII regions \cite[e.g] [] {Baccigalupi01}. In the frequency
range of interest ($\nu \la 10\GHz$), the two-point correlation function is
dominated by Galactic synchrotron emission, because of its significant
power on large angular scales. The observed brightness of the radio sky on
these scales, dominated by Galactic foreground, is discussed extensively in
\S \ref{sec:LFRB}. Here we focus on the level of contamination by Galactic
synchrotron emission in high Galactic latitudes and on small angular scales.

With large uncertainties regarding the spatial distribution of Galactic cosmic rays and magnetic fields, the best estimates of Galactic synchrotron emission are obtained by extrapolating direct measurements carried out at frequencies and angular scales where synchrotron emission is reliably measured. We thus make the key assumption, often used in the literature \cite[in particular in CMB anisotropy studies, e.g.][]{Tegmark00}, that the multipole dependence of the Galactic synchrotron APS varies little with frequency, such that $C_l(\nu) \simeq f(l) g(\nu)$. The frequency dependence $g(\nu)$ of Galactic synchrotron emission is discussed in \S \ref{sec:LFRB}. The APS multipole dependence $f(l)$ may be extracted from high resolution maps at $\sim\GHz$ frequencies down to arc-minute scales \cite[multipoles $l\simeq
6000$,][]{Tucci02}, and may in principle be extrapolated into the entire
frequency regime and multipole range of interest.

However, the angular power spectra extracted from radio maps depend on a number of factors, most importantly the Galactic latitude range examined and the efficiency at which discrete radio sources are removed from the map. Unfortunately, high resolution surveys are available mostly for low latitudes, which are less promising as potential search sites for an extragalactic signal because they exhibit a stronger Galactic foreground, in particular on small angular scales. Nonetheless, improved high and medium latitude data have recently made it possible to extend our knowledge of the high latitude APS up to multipoles $l\simeq 800$. 

The high latitude ($60.5\de<b<84.5\de$) low resolution (FWHM $0.6\de-2.3\de$) maps of Brouw \& Spoelstra (1976) for 5 frequencies between $408\MHz$ and $1411\MHz$, have been used to estimate the APS up to $l\simeq 70$ \cite[e.g.][table 6] {Bruscoli02}. Analysis of the Bonn $408\MHz$ full-sky survey \cite{Haslam82} suggests a power-law APS of the form $C_l(l\gg 1)\propto l^{-\beta}$ with $\beta \simeq 2.5-3.0$ down to the survey resolution limit $0.85\de$, corresponding to multipoles $l\simeq 200$ \cite{Tegmark96}. With a characteristic coherence scale $\sim 5\de-10\de$ for synchrotron emission \cite[e.g.][]{Spoelstra84,Banday91}, Tegmark \& Efstathiou (1996) argue that $C_l\propto (l+5)^{-3}$ describes the APS rather well. Thus, although the contribution of Galactic synchrotron emission to the two-point correlation function is significant, it is dominated by large angular scales around $5\de-10\de$ (low multipoles $l < 40$) and introduces \emph{little} contamination on small angular scales (see Figure \ref{fig:Sky_Cl}). The APS exhibits strong fluctuations across the sky, suggesting that 'quiet' regions may be identified and selected as preferable search sites for an extragalactic signal.  

Analysis of the $2.3 \GHz$ Rhodes map \cite{Jonas98}, with FWHM
resolution $20^\prime$, reveals strong variations of the APS with Galactic
latitude, and a large contribution of discrete radio sources which tend
to flatten the APS \cite{Giardino01}. At high latitudes ($|b|>20\de$),
Giardino et al. (2001) find, after removing discrete sources using median
filtering, that \beq C_l \simeq A^2 l^{-\beta} \coma \eeq with $A=0.3\pm
0.2\K$, and $\beta=2.92\pm0.07$, valid up to $l\simeq 100$. Although the
quoted uncertainties are large, comparison with the low resolution studies
mentioned above indicates that the normalization $A$ is probably no larger
than $0.3\K$. Similarly steep APS, with $\beta$ varying in the range
$2.60-3.35$ for different medium latitude ($|b|\la 20\de$) regions, were
found from a $1.4\GHz$ survey carried out with the Effelsberg $100$ m
telescope \cite{Uyaniker99}, after removing $>5\sigma$ unresolved discrete
sources \cite{Baccigalupi01}. The high angular resolution of the Effelsberg
survey, $\theta_c\sim 9.35^\prime$, thus suggests that the steep power-law
found by Giardino et al. holds up to multipoles $l\simeq 800$. Note that
after removing the brightest intensity peaks associated with low latitude
\HII regions, the APS highly resembles the APS of the polarized intensity
component \cite{Baccigalupi01}, as expected for synchrotron emission.

\subsubsection{Discrete Radio Sources}

Discrete radio sources, mostly radio galaxies, active galactic nuclei
(AGNs) and normal galaxies, make an important contribution to the
extragalactic radio background. Catalogues of radio sources at several
frequencies have been used to estimate their contribution to the radio sky
\cite{Simon77, Willis77}. This procedure is limited by the uncertain
contribution of faint, unidentified sources (mostly normal galaxies), which
dominate the background \cite[see for example][] {Windhorst93} and must be
modelled. Other studies have estimated the contribution of discrete sources
to the radio background, utilizing the well-known correlation between the
infra-red and radio flux densities of individual galaxies
\cite{Protheroe96,Haarsma98}. However, the results of such studies are
sensitive to the unknown redshift evolution of the sources. In addition,
the radio-infra-red correlation holds only in cases where the radio
emission is associated with star formation. Hence, such studies must be
supplemented by independent estimates of sources not associated with star
formation, such as a catalogue-based estimate of the emission from AGNs
\cite{Ryle68}.

\begin{figure*}[h]
\epsscale{1.2}
\plotone{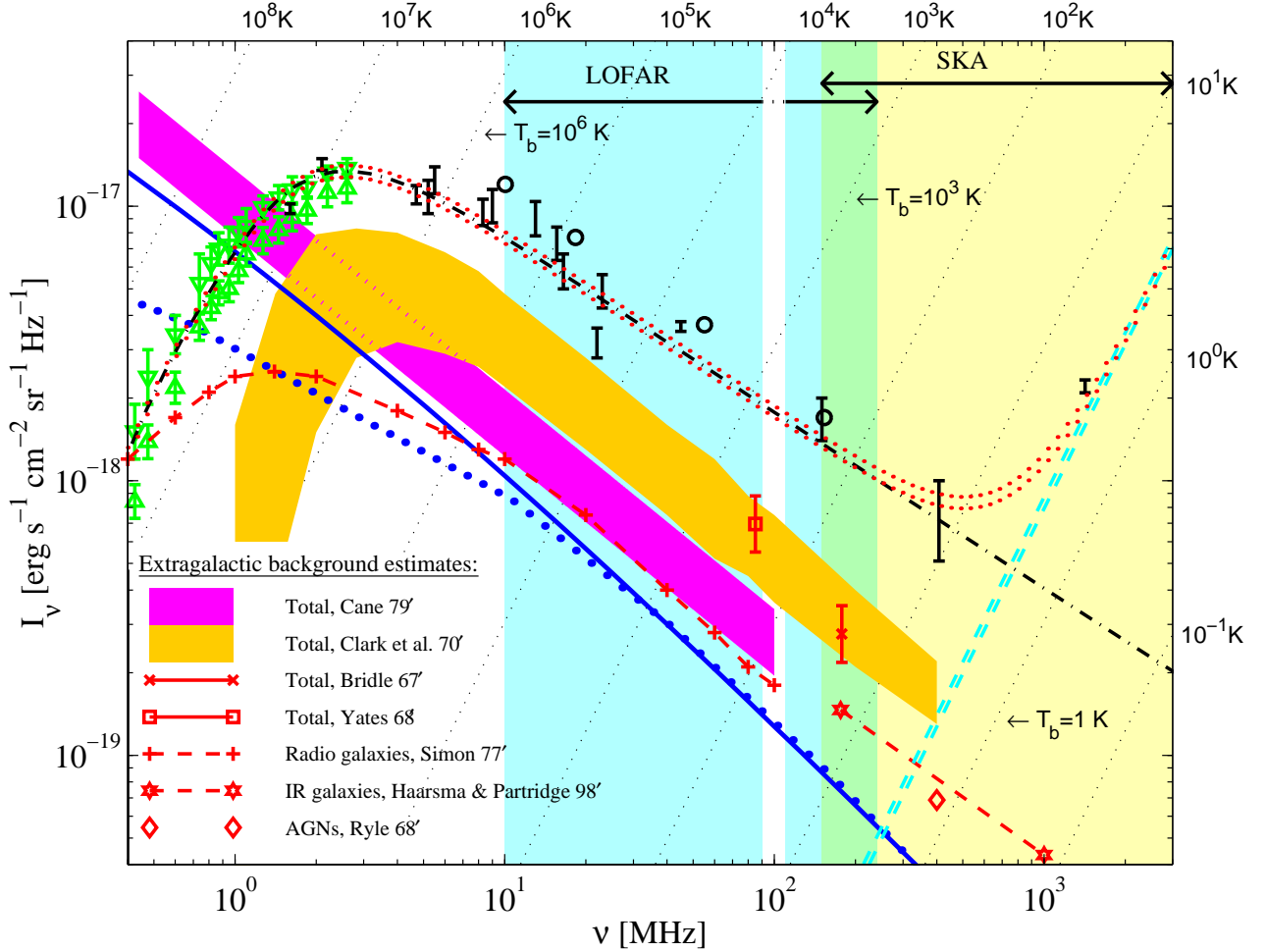}
\caption{ Brightness of the radio sky. Ground-based (circles and error bars for the Galactic polar regions) and space-based (triangles for IMP-6 spacecraft minimum and maximum) observations are dominated by Galactic synchrotron emission (dash dotted line shows the modelled Galactic foreground toward the polar regions, see \S \ref{sec:LFRB}) in low frequencies, and by the CMB (double-dashed line) at high frequencies. Various estimates of the extragalactic component (legend, see text) are typically an order of magnitude lower than the Galactic foreground. The background from intergalactic shocks according to the calibrated model (of \S \ref{subsec:model_LCDM_sim} and Table \ref{tab:ModelParams}, solid line) is roughly of the same magnitude, and scales according to $I_\nu\propto f_{acc} f_T^2 \widetilde{f}_r^{-2}$. The signal is also shown according to the \LCDM simulation \cite[accounting only for emission above the spectral break, dotted line, see] [] {Keshet04b}. The frequency ranges of the LOFAR and the SKA are shown (double arrows and corresponding shaded regions), as well as constant brightness temperature contours (dotted lines, labelled on the top and on the right axes).  }
\label{fig:LFRB_sky}
\end{figure*}

The studies outlined above suggest that the contribution of discrete sources to the radio sky is roughly an order of magnitude lower than the observed sky brightness. This corresponds to $25\%-65\%$ of the total extragalactic background, as estimated by spectral modelling of low frequency radio observations \cite[][see \S \ref{sec:LFRB}]{Clark70,Cane79}, although some evolutionary models \cite[e.g.] [] {Protheroe96} may even account for the entire strong extragalactic signal calculated by Clark et al. (1970). The spectral index of the integrated emission from discrete sources has been estimated to lie in the range $s=0.7-0.8$ for frequencies in the range $100\MHz-\mbox{few} \GHz$ \cite{Simon77,Lawson87,Haarsma98}, although some dependence upon scale may be expected because the source distribution is essentially bimodal \cite{Tegmark00}. At very low ($\nu\la10\MHz$) frequencies, the background from discrete sources turns around, probably because of synchrotron self-absorption \cite{Simon77}. 

It is easier to estimate the contribution of discrete sources to the APS,
because the latter is dominated by the bright, well studied
sources. Discrete radio sources have approximately a Poisson distribution
in the sky, because projection through their wide redshift distribution
effectively diminishes their correlations \cite [e.g.][]
{Tegmark96}. Hence, as long as the angular scales concerned are much larger
than the angular extent of the sources, the two-point correlation function
vanishes and the APS is flat, $C_l\propto l^0$ (white noise), such that the logarithmic contribution to the variance is $\delta I_l \propto l$. One may estimate the APS as \cite[e.g.][]{Tegmark96}
\beq \label{eq:Cl_Poisson} C_l = \int_0^\infty S^2 {\partial N \over
\partial S} \,dS \coma \eeq where $S$ is the source flux and $\partial
N/\partial S$ is the differential number density of discrete sources in the sky. The limited sensitivity $S_{min}$ of the observations limits our knowledge of $\partial N/\partial S$ to sources with flux $S>S_{min}$, but since the number density of faint sources is not too steep, the possible error introduced is small. The upper limit of the integral, essentially determined by the brightest sources, can be lowered in order to reduce the noise, by modelling and removing the brightest sources from the analyzed map. However, this quickly becomes laborious as the number of sources increases, and introduces inevitable errors associated with source removal uncertainties.

\begin{figure*}[h]
\plotone{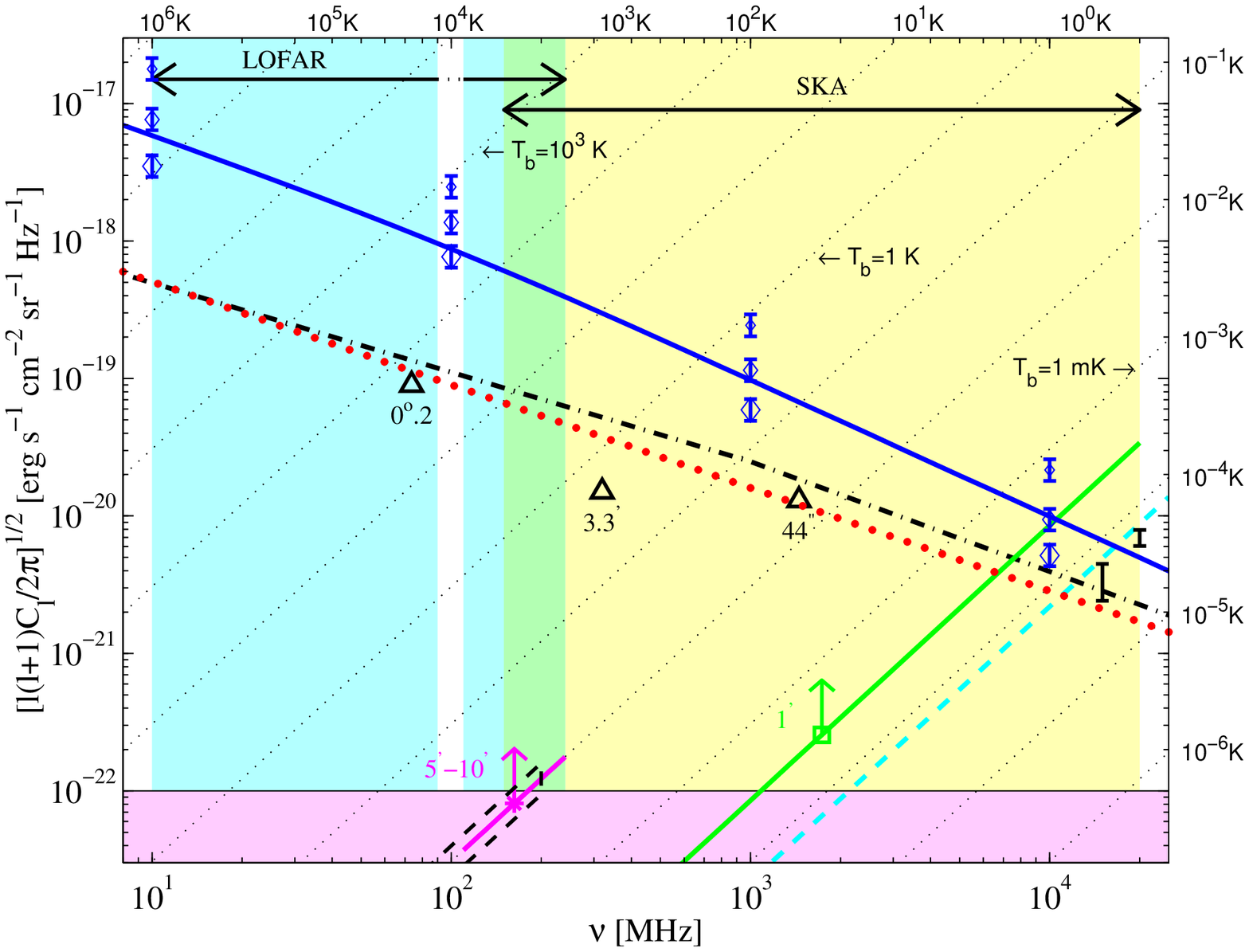}
\epsscale{1.2}
\caption{ Logarithmic contribution $\delta I_l$ of various signals to the variance at multipole $l=400$ (corresponding to $\theta\simeq 0\de.5$). The background from intergalactic shocks is shown according to the calibrated model (of \S \ref{subsec:model_LCDM_sim} and Table \ref{tab:ModelParams}, solid line), and scales roughly according to $\delta I_l(l; \, \xi_e , \xi_B , f_{acc},f_T,\widetilde{f}_{r}) = \xi_e \xi_B f_{acc} f_T^2 \widetilde{f}_{r}^{-3} \, \delta I_l(\widetilde{f}_{r}l;\,1,1,1,1,1)$. The signal according to the \LCDM simulation \cite[diamonds with statistical error bars for $l=400$, $10^3$ and $10^4$, from bottom to top, accounting only for emission above the spectral break, see] [] {Keshet04b} scales linearly with $\xi_e \xi_B$. The high-latitude Galactic foreground (dash-dotted line) and the integrated emission from discrete radio sources (dotted line, assuming removal of sources above $S_{cut}=100\mJy$ at $1.4\GHz$) are presented. At frequencies above $\sim10\GHz$, CMB fluctuations take over \cite[dashed line extrapolated from WMAP measurements, error bars present detections by the Cambridge Radio Telescope on $0\de.2-0\de.5$ scales, and by the Owens Valley Radio Observatory on $0\de.1-0\de.6$ scales; for references see][] {Hinshaw03, White99}. Bremsstrahlung emission from \Lya clouds (low horizontal shaded region) and the maximal 21 cm tomography signal (dashed contour) are seen to be much weaker than the above competing signals. Continuum surface brightness sensitivities of the LOFAR (star) and the SKA (square), and the side-lobe confusion limit of the VLA (triangles, for 10 minutes in configuration D, see http://www.vla.nrao.edu/astro/guides/vlas) are shown (labels denote beam widths). Constant brightness temperature contours are shown as dotted lines, labelled on the top and the right axes. Emission from intergalactic shocks is seen to dominate the sky at this angular scale for frequencies $\nu \la 500\MHz$, provided that $\xi_e \, \xi_B \ga 10^{-4}$. According to the simulation, the emission dominates the sky at this frequency range for $\theta <0\de.5$, even if $\xi_e \, \xi_B = 5\times 10^{-5}$.  }
\label{fig:Sky_Cl}
\end{figure*}

Following Tegmark \& Efstathiou (1996), we use source counts produced by
the $1.4\GHz$ VLA FIRST all-sky survey. A small fraction ($\sim 10^{-4}$)
of the discrete sources in the FIRST catalogue have angular scales larger
than an arc-minute \cite{White97}, so the use of equation
(\ref{eq:Cl_Poisson}) is legitimate when dealing with multipoles $l\la
10^4$. The number density of radio sources \cite[after performing a
resolution correction for extended sources,][]{White97} is well fit in the
flux range $S=1\mJy-1\Jy$ by 
\beq {\partial N \over \partial S} \simeq 1.5\times 10^8 \left(S \over 1\mJy \right)^{-1.5} \left(1 + {S \over 100 \mJy} \right)^{-1} \Jy^{-1} \sr^{-1} \fin \eeq 
For faint sources in the range $S=10\muJy-1\mJy$, the number count steepens to $\partial N /\partial S \propto S^{-\gamma}$ with $\gamma = 2.2\pm 0.2$ \cite{Windhorst93}. This changes the integrated brightness considerably, but has only a negligible effect on the APS. Using equation (\ref{eq:Cl_Poisson}), we find that for a modest cut at $S_{max}=100\mJy$, corresponding to removal of the $\sim 7\times 10^4$ brightest sources all-sky, $C_l(\nu) \simeq 1.7 \times 10^{-8} (\nu/1.4\GHz)^{-2.75} \K^2$. Using the 6C survey carried out at $151\MHz$ \cite{Hales88}, we find that the above cut is equivalent to
removal of sources brighter than $\sim 3\Jy$ at $150\MHz$. As shown in Figure
\ref{fig:Sky_Cl}, such a cut places the emission from discrete sources at the
same level as the Galactic foreground for angular scales $\sim 0\de.5$. At
smaller angular scales, contamination from discrete sources becomes
increasingly worse, requiring the removal of more sources.

\subsubsection{Bremsstrahlung Emission from \Lya Clouds}
\label{sec:brehmsstrahlung}
\indent

Loeb (1996) had calculated the thermal bremsstrahlung emission from \Lya clouds. These clouds are most likely in photo-ionization equilibrium with the UV background, permitting an estimate of the bremsstrahlung signal if the UV background intensity, $J_\nu^{uv}(z)$, is known. In the following, we update the predicted bremsstrahlung signal according to recent UV observations. 

Based on the neutral hydrogen density estimated from $\mbox{Ly}\alpha$ absorption lines in the spectra of background QSOs, the emission from \Lya clouds with column densities below the Lyman limit ($N \la 10^{17} \mbox{ cm}^{-2}$) at a redshift $z<5$, is given by:
\beq J_\nu \simeq 10^{-22.5\pm0.4} \langle J_{22} \rangle \IUnits \fin \eeq 
Here $\langle J_{22} \rangle$ was defined as the UV intensity above the Lyman limit (wavelengths $\lambda < 912$\AA) weighted over its redshift evolution
\beq \langle J_{22} \rangle \equiv \frac {\int_0^5 J_{22}(z) (1+z)^{-(3-\gamma)} dz}{\int_0^5 (1+z)^{-(3-\gamma)} dz} \coma \eeq
where $J_{22}\equiv J_\nu^{uv}(z) / (10^{-22} \IUnits)$, and the fit parameter $\gamma =1.5\pm 0.4$ is determined by the redshift dependence of the \HI cloud surface density \cite[see][and the references therein] {Loeb96}. The signal is flat for frequencies $\nu \ll 10^{14}\Hz$, persisting throughout the frequency range of interest.  

The UV background intensity may be estimated using the proximity effect, whereby one compares the effects of photo-ionization by the UV background and by a nearby QSO on a \Lya cloud. Scott et al. (2002) have thus estimated the UV background in redshifts $z < 5$ as follows:
\beq J_{22}(z) \label{eq:J22} \simeq \cases{ 
0.65_{-0.45}^{+3.8}  &  for $z<1$ ; \cr
1.0_{-0.22}^{+3.8}   &  for $1<z<1.7$ ; \cr
7.0_{-4.4}^{+3.4}    &  for $1.7<z<5$ , \cr} \eeq
implying that 
\beq \label{eq:total_brem_flux} J_\nu = 10^{-22 \pm 1} \IUnits \fin \eeq 
This result neglects the contribution of optically thick clouds at high ($z>5$) redshifts, and possible additional ionization fields. Hence, the actual bremsstrahlung flux may be somewhat higher than calculated in equation (\ref{eq:total_brem_flux}). 

Loeb (1996) had also evaluated the amplitude of intensity fluctuations introduced by the finite number of \HI clouds along any given line of sight. Assuming a random cloud distribution in redshift and in column density, one finds 
\beq \label{eq:brem_rms} \Delta J_\nu = 10^{-23\pm0.4} \langle J_{22}^2 \rangle^{1/2} \IUnits \coma \eeq
where we have defined 
\beq \langle J_{22}^2 \rangle \equiv \frac {\int_0^5 J_{22}(z)^2 (1+z)^{-(6-\gamma)} dz}{\int_0^5 (1+z)^{-(6-\gamma)} dz} \fin \eeq 
Using equation (\ref{eq:J22}), we thus find $\Delta J_\nu = 10^{-23\pm1} \IUnits$. Such fluctuations are expected on $\sim 7^{\prime \prime}$ scales (corresponding to $l\simeq 10^5$), although weaker fluctuations introduced by clouds associated with large-scale structure may appear even on $\sim 10^\prime$ scales. This fluctuation level should be regarded as a lower limit, because in addition to the assumptions leading to equation (\ref{eq:total_brem_flux}), the non-randomness of the cloud distribution further enhances the fluctuation signal. Nevertheless, without a unique spectral or temporal signature, confusion with the other signals discussed in this section, in particular discrete sources, is likely to preclude detection of bremsstrahlung from \Lya clouds throughout the frequency range of interest.

\subsubsection{IGM 21 cm Tomography}
\label{sec:21tomography}
\indent

Madau, Meiksin \& Rees \cite[1997, see also][]{Tozzi00} have proposed that the 21 cm ($1.4\GHz$) spin flip transition of atomic hydrogen could be used to probe the IGM at the epoch prior to reionization. Spatial inhomogeneities in the IGM may be observed today as redshifted emission or absorption fluctuations against the CMB, in situations where early sources of radiation decoupled the IGM spin temperature from the CMB temperature. In principle, the combined angular and spectral signal, stronger than the CMB fluctuations by two orders of magnitude, may be used to trace the 'cosmic-web' structure of the early universe, in both space and cosmic time. 

However, as pointed out by Di Matteo et al. (2002), at the relevant
frequency range $50-200\MHz$, contamination by discrete radio sources
imposes a serious contamination even for the maximal signal amplitude
$\Delta T\simeq 10\mK$ expected. Nevertheless, a \emph{spectral}
$\sim10\mK$ feature, caused by the fast rise of the \Lya background as the
first UV sources reionized the IGM, could possibly be detected behind the
spectrally continuous foreground (Zaldarriaga, Furlanetto, \& Hernquist
2003; Morales \& Hewitt 2003; Loeb \& Zaldarriaga 2003; Gnedin \& Shaver
2003, and references therein). Emission from intergalactic shocks provides
an additional important source of confusion for $21\cm$ tomography, but its
smooth spectrum may not prevent detection of the sharp 21cm spectral features.

\subsection{Summary and Conclusions} 
\label{subsec:feasibility_summary}
\indent

We conclude that the design of next generation radio telescopes such as the LOFAR, the SKA and the ALFA, is more than sufficient for detection of the angular fluctuations introduced by intergalactic shocks, as calculated in \S \ref{sec:IGM_model} and confirmed by a cosmological simulation \cite{Keshet04b}. Identification of the signal is limited by confusion with Galactic foreground and with discrete radio sources. Foreground fluctuations in the synchrotron emission of our Galaxy constrain a clear detection of the signal to sub-degree scales and to high Galactic latitudes. Confusion with discrete radio sources requires that the brightest sources be modelled and removed. With a feasible point source cut ($100\mJy$ at $1.4 \GHz$, or $3\Jy$ at $150\MHz$), the signal dominates over the competing signals at angular scales $10\arcmin \la \theta \la 1\de$ (by roughly an order of magnitude according to the \LCDM simulation), whereas detection on arcminute scales will require a more ambitious point source removal. The spectrum of emission from intergalactic shocks indicates that the signal is most pronounced at $\sim 100\MHz$ frequencies, planned to be covered by both the LOFAR and the SKA. We find that other extragalactic signals in the radio band, namely bremsstrahlung from \Lya clouds and angular (but perhaps not spectral) 21 cm tomography, will be confusion limited by intergalactic shocks as well as by radio point sources.

Interestingly, detection of the fluctuating signal from intergalactic shocks is just possible with \emph{present} high resolution radio telescopes, such as the VLA (at the maximal sensitivity configuration, see Figure \ref{fig:Sky_Cl}). The calculated signal should be detectable at high latitudes, a few $100\MHz$ frequencies and sub-degree scales. As noted by Waxman \& Loeb (2000), the signal may have already been detected by CMB anisotropy studies, at $\sim 10\GHz$ frequencies and $\theta <1\de$ scales. In particular, the signal may have been detected in arcminute scales by low frequency CMB anisotropy experiments such as the Australian Telescope Compact Array (ATCA), Ryle, and the VLA. It is possible, that some of the noise removed from the corresponding maps \cite[e.g.] [] {Subrahmanyan2000} is associated with emission from intergalactic shocks, and is thus correlated with tracers of large-scale structure. 

The high resolution of present $\sim \GHz$ high resolution surveys raises
the possibility that their maps already include the signal at an
identifiable level. Such surveys are mostly available at low latitudes
(say, $|b|<8\de$), and thus exhibit stronger Galactic foreground with
flatter angular power spectra than found in high latitudes. This implies
significant foreground contamination, especially on small angular
scales. Some low resolution studies have found an APS power index
$\beta\simeq 2.4$ extending up to $l\simeq 900$ \cite{Giardino01,
Tegmark00}, others finding $\beta\simeq 1.7$ extending up to $l\simeq 6000$
\cite{Tucci02}. In addition, detection of the signal becomes increasingly difficult for frequencies away from the optimal frequency, $\sim 100\MHz$. Nevertheless, large longitudinal fluctuations have been identified in the low latitude APS \cite[e.g.] []{Baccigalupi01}, suggesting that the signal may surface in the 'quiet regions'. Analysis of such regions, including careful removal of discrete sources, may yield the desired signal, identifiable at $\sim 10^\prime$ scales by cross-correlating the maps with known tracers of large scale structure.


\section{Analysis of the Low Frequency Radio Sky}
\label{sec:LFRB}

Next, we analyze the diffuse low frequency ($\nu <500 \MHz$) radio background (LFRB), dominated by Galactic synchrotron emission. A simple model for the Galactic emission is presented in order to (i) try to separate between the Galactic foreground and the extragalactic background; (ii) examine if a simple Galactic model can account for the observed Galactic foreground; and (iii) demonstrate the importance of observations in the frequency range $1\MHz \la \nu \la 10\MHz$, where free-free absorption in our Galaxy is significant. 

We find that the observations are well-fit by a simple double-disk Galactic model, given the existing observational uncertainties. This implies that the uncertainties and low resolution of present low frequency observations \emph{preclude} a direct identification of the diffuse extragalactic radio background (DERB). Upper limit on the DERB flux at frequencies $\nu\la 30$ MHz may be imposed, at a level roughly an order of magnitude lower than the Galactic foreground. Point sources and intergalactic shocks already account for a substantial fraction of this upper limit, suggesting that it is not far from the true extragalactic signal.

We point out the lack of models accounting for the \emph{combined} spectral and angular data at very low ($1\MHz\la\nu\la10\MHz$) frequencies, where absorption in our Galaxy is non-negligible. We discuss the frequency-dependent anisotropy pattern observed in this frequency range, reflecting different emission and absorption along different lines of sight. We demonstrate, when constructing the Galactic model, how the combined spectral and angular data may be used to disentangle the distributions of Galactic cosmic-rays, magnetic fields and ionized gas, leading to an elaborate three-dimensional model of the Galaxy. Future high resolution observations at very low frequencies, produced for example by the ALFA space mission, could thus provide valuable information regarding the structure and composition of the Galaxy.  

We begin by reviewing various spectral and angular features of LFRB observations, in \S\ref{subsec:LFRB_observations}. A simple model for Galactic emission and absorption is presented in \S\ref{subsec:LFRB_model}, and shown to fit the combined spectral and angular data rather well. Previous LFRB models, which in general account for either the spectral or the angular aspects of the observations, are reviewed in \S \ref{subsec:LFRB_prev_models}, and claims for direct identification of the DERB are discussed.

\subsection{Observations of the Low Frequency Radio Background (LFRB)}
\label{subsec:LFRB_observations}

The LFRB has by now been measured by a variety of ground-based and
space-borne instruments.  Detailed maps of the low frequency radio sky have
been produced by ground-based experiments and published for frequencies
$1.6\MHz-16.5\MHz$ \cite{Ellis82, Ellis87}, $22\MHz$ \cite{Roger99},
$34.5\MHz$ \cite{Dwarakanath90}, $38\MHz$ \cite{Milogradov73}, $45\MHz$
\cite{Alvarez97}, and $408\MHz$ \cite{Haslam82}.  The radio background in
frequencies below $5$ MHz is inaccessible to most ground-based experiments,
because of strong absorption by the ionosphere.  Notable exceptions are the
radio telescopes operated in Tasmania, taking advantage of the unusually
favorable ionospheric conditions prevailing in the region \cite{Ellis65} to
measure the background in frequencies as low as $1.6$ MHz \cite{Ellis87}.
At even lower frequencies, one must resort to space telescopes such as the
Radio Astronomy Explorer 1 (RAE-1) satellite \cite{Clark70,Alexander70},
the radio astronomy experiment on board the IMP-6 spacecraft
\cite{Brown73}, the RAE-2 lunar orbiting satellite \cite{Novaco78} and the
WAVES experiment on board the WIND spacecraft \cite{Manning2001}. Our
picture of the very low frequency radio sky will be revolutionized by
future high resolution space telescopes, such as the ALFA mission, planned
to operate at $30\kHz-30\MHz$ frequencies (see Table
\ref{tab:TelescopeParams}).

\begin{figure}[b]
\epsscale{1.2}
\plotone{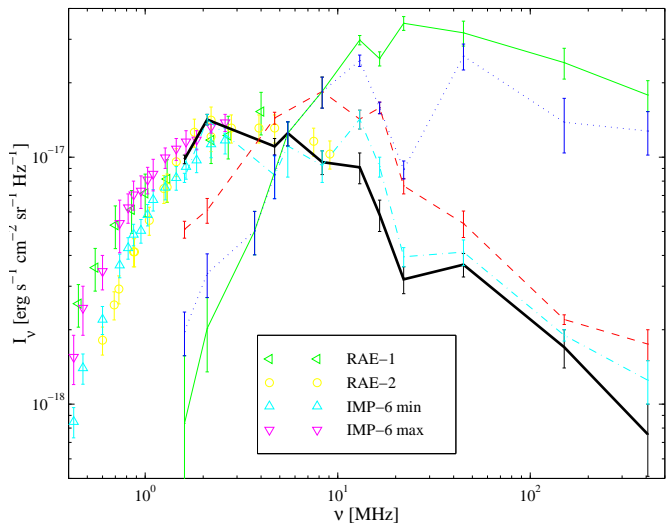}
\caption{ Low frequency observations. Ground-based data (error bars, lines
are a guide to the eye) are shown for five regions in the sky (see text and
Table \ref{tab:skymap}): the Galactic center [GC, $(l,b)=(0,0)$,
\emph{solid line}]; near the GC [NGC, ($20\de,0$), \emph{dotted line}];
near the anti-center [NAC, ($240\de,0$), \emph{dashed line}]; a south polar
region [SPR, ($30\de,-70\de$), \emph{heavy line}]; and a low brightness
region near loop I [LI, ($320\de,50\de$), \emph{dash-dotted line}]. In
spite of the large statistical and systematical errors (involving poor and 
frequency dependent angular resolutions, temperature scales, and
zero-point temperatures), the frequency-dependent anisotropy pattern is
apparent. The low resolution satellite data (see legend) approximately
reflect the brightness at high Galactic latitudes, which dominate the
sky at very low frequencies.  }
\label{fig:LFRB_raw}
\end{figure}

Generally, the ground-based experiments achieve better angular resolution, ranging from sub-degree scales at high frequencies (e.g. $51^\prime$ at $408$ MHz) to $\ga10\de$ at low ($\sim 2$ MHz) frequencies, whereas space-borne telescopes suffer from very poor angular resolution, of order $\sim 100\de$. On the other hand, ground-based experiments are generally limited not only to higher frequencies, but also to smaller fields of view, requiring a combination of several long-term surveys in order to produce an all-sky map, such as the Bonn all-sky radio continuum survey \cite{Haslam82}. Figure \ref{fig:LFRB_raw} illustrates various low frequency observations, carried out by both ground-based and space-borne telescopes. Ground-based observations were extracted from published sky maps, and are quoted for several interesting regions on the sky for which multi-frequency data exist, such as the Galactic center (GC) and the south Galactic pole region (SPR). The data are summarized in Table \ref{tab:skymap}.

\begin{deluxetable*}{llllllll}
\tablecaption{\label{tab:skymap} Summary of low frequency ground-based radio observations\tablenotemark{a}}
\tablewidth{0pt}
\tablehead{ \em $\nu$ \tablenotemark{b} & $R$ \tablenotemark{b} & \em SP \tablenotemark{c} & SPR \tablenotemark{c} & \em GC \tablenotemark{c} & \em NGC \tablenotemark{c} & \em NAC \tablenotemark{c} & \em LI \tablenotemark{c} }
\startdata
1.6  & 25 & 8.6-9.4 & 9.4-10.2 & 0-1.57 & 1.57-2.36 & 4.7-5.5 & 7.9-8.6 \\ 
2.1  & 7.5 & \nodata & 13.5-14.9 & 1.35-2.70 & 2.70-4.06 & 5.4-6.8 & 13.5-14.9 \\ 
4.7  & 6.8 & \nodata & 10.2-11.9 & 4.0-6.0 & 6.78-10.2 & 13.6-15.2 & 6.8-10.1 \\ 
5.5  & 4.5 & 5.6-8.3 & 11.1-13.9 & 11.1-13.9 & 11.1-13.9 & \nodata & 8.3-13.9 \\
8.3  & 3.0 & \nodata & 8.5-10.6 & 15.8-21.1 & 15.8-21.1 & 15.8-21.1  & 7.9-10.6 \\ 
13.0 & 1.9 & 5.2-7.8 & 7.8-10.4 & 28.5-31.1 & 23.3-25.9 & 13.0-15.5 & 13.0-15.5 \\ 
16.5 & 1.5 & \nodata & 5.0-6.7 & 23.4-26.7 & 15.0-16.7 & 15.0-16.7 & 8.3-10.0 \\ 
22.0 & 1.7 & 3.6-4.3 & 2.8-3.6 & 32.6-37.1 & 8.16-9.65 & 7.1-8.2 & 3.6-4.3 \\
45.0 & 4.6 & 3.3-3.4 & 3.6-3.7 & 31.0-32.3 & 24.8-26.1 & 5.2-5.5 & 4.0-4.2 \\
150  & 2.2 & 1.4-2.0 & 1.4-2.0 & 20.7-27.6 & 10.4-17.3 & 2.1-2.3 & 1.8-2.0 \\
408  & 0.85 & 0.51-1.0 & 0.51-1.0 & 15.3-20.4 & 10.2-15.3 & 1.5-2.0 & 1.0-1.5 \\
\enddata
\tablenotetext{a}
{ Data extracted from maps produced by Ellis \& Mendillo 1987 (1.6 MHz), 
Ellis 1982 (2.1-16.5 MHz), Roger et al. 1999 (22 MHz), Alvarez 1997 (45 MHz), 
Landecker \& Wielebinski 1970 (150 MHz), and Haslam et al. 1982 (408 MHz). }
\tablenotetext{b}
{ Symbols: $\nu$ - frequency (MHz), $R$ - angular resolution (degree).}
\tablenotetext{c}
{ Brightness (in $10^{-18}\IUnits$) measured toward various parts of the sky for which multi-frequency data exist: 
SP - the south Galactic pole ($l,b$)=$(0\de,-90\de)$; 
SPR - a south polar region $(30\de,-70\de)$; 
GC - the Galactic center ($l,b$)=$(0\de,0\de)$;
NGC - the Galactic plane, near the Galactic center $(20\de,0\de)$;
NAC - the Galactic plane, near the Galactic anti-center $(240\de,0\de)$;
and LI - a region of low brightness within loop I $(320\de,50\de)$.}
\end{deluxetable*}

\subsubsection{Spectra}

The low frequency radio spectrum is often approximated as a broken power-law. The specific intensity is found to peak at frequencies in the range $\nu_{peak} \simeq 2-20\MHz$, where $\nu_{peak}$ varies across the sky (Figure \ref{fig:LFRB_raw}). 
In the frequency range $20-200$ MHz, the specific intensity $I_\nu$ measured towards a given line of sight decreases with increasing frequency, and is well fitted by a power-law $I_\nu \propto \nu^{-s}$, with spectral indices in the range $s\simeq 0.2-0.7$ varying across the sky. For example, $s=0.55\pm 0.03$ is found toward the north Galactic pole, $s=0.65\pm0.15$ toward the south Galactic pole, $s=0.38 \pm 0.02$ toward the Galactic anti-center \cite[see Figure \ref{fig:LFRB_raw}, ][and the references therein]{Bridle67,Cane79} and $s\sim 0.25$ towards the Galactic center. At low frequencies ($\nu<\nu_{peak}$), on the other hand, $I_\nu$ increases as a function of increasing frequency. For example, the specific intensity measured toward the Galactic polar region rapidly drops below $3-5$ MHz (where angular resolution is exceedingly poor), falling by an order of magnitude as the frequency is lowered to $\sim 400$ kHz, and by another order of magnitude by $\sim 250$ kHz \cite{Brown73, Dulk01}.

At relatively high frequencies, around $200-400$ MHz, the spectrum steepens from spectral indices in the range $0.2-0.7$ to $s \simeq 0.8-0.9$, with small deviations in the sky, the largest of which are associated with the Galactic spurs \cite[loops I, III and IV, which exhibit spectra flatter in their center and steeper in their ridges, see][]{Bridle67,Webster74}. At very low frequencies, below $300\kHz$, an excess radiation has been observed \cite{Brown73,Manning2001}, flattening the spectrum. Although the excess was previously accounted for by a hot, tenuous ISM phase \cite{Fleishmann95}, recent observations indicate that the excess radiation is localized \cite[at ecliptic coordinates $l_{ecl}\simeq 30\de$ or $l_{ecl}\simeq 210\de$, see][]{Manning2001}, suggesting that the radiation originates from a local source. We shall not deal with low frequencies below $400\kHz$ in the following discussion.

\subsubsection{Anisotropy Pattern}

The LFRB is in general \emph{anisotropic}, the radio maps revealing structure on various scales. Structure on hemispheric scales is evident, indicating that most of the radiation is Galactic, and thus reflecting the structure of the Galaxy and the position of the solar system within it. The anisotropy pattern of the LFRB is highly frequency dependent. Figure \ref{fig:LFRB_raw} illustrates this anisotropy pattern, by comparing the spectra measured towards different directions in the sky. 

All experiments indicate that for frequencies $\nu \ga 6\MHz$, the
background is strongest toward the Galactic plane, the brightest emission
originating from the direction of the Galactic center itself.  In the
frequency range $600\mbox{ kHz}<\nu<2.5\mbox{ MHz}$, on the other hand, the
radiation is dominated by emission originating from the Galactic polar
regions, according to most of the experiments sensitive to this energy
range: measurements by the RAE-1 satellite indicate that the radiation from
the polar region is stronger than the radiation from the Galactic center
and anti-center below $\sim 1 \mbox{ MHz}$ \cite{Clark70,Alexander70}; the
IMP-6 experiment suggests with a high probability that the radiation
between $130$ kHz and $2.6$ MHz is maximal towards the Galactic poles and
minimal towards the ecliptic plane \cite{Brown73}; and the WAVES experiment
indicates that the radiation is dominated by emission from the Galactic
polar regions between $\sim 600$ kHz and $\sim 2.5$ MHz, is essentially
isotropic at $\sim 3.5$ MHz, and is dominated by emission from the Galactic
center above $\sim 5$ MHz \cite{Manning2001}.  On the other hand, it is
interesting to note that observations carried out by the RAE-2 spacecraft
find no such phenomenon and indicate domination of the Galactic disk over
the Galactic polar regions at all frequencies above $\sim 500$ kHz
\cite{Novaco78}.  Most illuminating are the sky maps obtained by very low
frequency ($1.6-5.5\MHz$) ground-based telescopes positioned in Tasmania
\cite{Ellis82,Ellis87}, which suggest a transition in the anisotropy
pattern taking place between $2.1\MHz$ and $3.7\MHz$; whereas above $3.7
\MHz$ the emission is stronger toward low latitudes (near the Galactic
plane), the pattern below $2.1$ is reversed, stronger emission detected
toward high latitudes (near the Galactic poles). This effect, although
small, is significant and consistent with most space-based observations
carried out at low frequencies and with all ground-based observations at
high frequencies.

\subsection{Galactic Model}
\label{subsec:LFRB_model}

Since the 1950s, it has been widely accepted that the LFRB is dominated by
Galactic synchrotron radiation that is emitted as relativistic (cosmic-ray)
electrons gyrate in the magnetic fields reposing the ISM. At higher
frequencies, the synchrotron signal is overtaken by competing radiative
processes, namely the CMB (above $1\GHz$), thermal bremsstrahlung
\cite[dominant above several GHz, see e.g.][]{Lawson87}, and emission from
spinning dust \cite[dominant above $\sim70\GHz$, e.g.] [] {Tegmark00, Finkbeiner03}. Transition radiation, emitted when charged particles
are accelerated in a medium with a varying refractive index, dominates at
very low frequencies \cite[$\nu\ll 100\kHz$,][]{Fleishmann95}. An
additional, extragalactic radio component must also be present, at least
partly attributed to the integrated emission from discrete radio sources \cite[e.g.][and the references therein]{Simon77} and to intergalactic shocks.

In this section we construct a simple Galactic model, based on important features of the observed LFRB (see \S \ref{subsec:LFRB_observations} and Figure \ref{fig:LFRB_raw}) and the present understanding of the Galactic distributions of cosmic-rays, ionized gas and magnetic fields. Large observational uncertainties and poor resolution render all but the simplest qualitative model redundant, but the latter is sufficient to show that the observed background may, in principle, be entirely Galactic. The good fit of our Galactic model to existing observations (within present uncertainties) shows that the introduction of an additional, extragalactic component is not \emph{necessary} for explaining the observations.

\subsubsection{Construction}

We begin by considering an over-simplified model, in which the Galactic composition is assumed uniform along a given line of sight; more specifically, we assume that the emissivity $j(\nu)$ and the absorption coefficient $\alpha(\nu)$ are roughly constant throughout the Galaxy, along the line of sight. Supplementing the Galactic emission by an additional extragalactic component of specific intensity $I_x(\nu)$, thus leads to a simple two-component model. The equation of radiative transfer may then be integrated to give the specific intensity as a function of the distance $L$ transversed along the line of sight through the Galaxy:
\beq \label{eq:two_comp} I(\nu,L) = j(\nu)\, L\, \frac{1-e^{-\tau(\nu)}}{\tau(\nu)}
 + I_x(\nu) e^{-\tau(\nu)} \mbox{ , }\eeq 
where $\tau(\nu) = \alpha(\nu) L$ is the optical depth. In the limits of very small or very large optical depths, this expression simplifies to:
\beq \label{eq:two_comp_limits} I(\nu,L) \simeq \cases{ 
j(\nu) L + I_x(\nu) & for $\tau(\nu)\ll 1$ ; \cr
j(\nu)/\alpha(\nu) + I_x(\nu) e^{-\tau(\nu)} & for $\tau(\nu)\gg 1$ .\cr } \eeq
Note that such a model implies that the observed intensity is a monotonically rising function of $L$ for frequencies where $I_x(\nu)<j(\nu)/\alpha(\nu)$ and is nearly isotropic in low frequencies, in contrast to the anisotropy pattern observed. In addition, uniform emission and absorption can not explain the observed sharp cutoff at frequencies $\nu\la800\kHz$ \cite[although for certain ISM parameters the cutoff may be attributed to the Razin-Tsytovich effect, see e.g.][and the references therein]{Fleishmann95}. 

The observed low-frequency cutoff at low Galactic latitudes scales roughly as $I_\nu \propto \nu^{-s}$ with $s\simeq -2$ (see Figure \ref{fig:LFRB_raw}), characteristic of mixed emission and absorption at a high optical depth [see equation (\ref{eq:two_comp_limits})]. On the other hand, at high latitudes the cutoff is stronger, $s\la-2.7$ for $\nu\la800\kHz$ (in contrast, for $\nu\ga20\MHz$ the spectrum is typically \emph{softer} at high latitudes). Such a cutoff is naturally explained by the presence of a thin layer of relatively dense, strongly absorbing ionized gas near the Galactic plane. A double-component Galactic model, consisting of a thin, dense disk and a thick, dilute disk (as found by previous models which have focused on the high frequency angular data, see \S \ref{subsec:LFRB_prev_models}), may also naturally explain the observed frequency-dependent anisotropy pattern of the LFRB. Unlike a single-component Galactic model (with or without an additional isotropic, extragalactic component), in such a model the brightness need not be a monotonic function of the distance $L$ transversed through the Galaxy along the line of sight [for frequencies where $I_x(\nu) < j(\nu)/\alpha(\nu)$]. 

Finally, the spectrum measured at a given direction is often well fit at low ($\nu\la$ few MHz) and at high ($\nu\ga$ few $10\MHz$) frequencies by equation (\ref{eq:two_comp_limits}). Hence, one may directly extract from the data the average values of $j(\nu)/\alpha(\nu)$ and $j(\nu)L$ towards various lines of sight [note that equations (\ref{eq:two_comp}) and (\ref{eq:two_comp_limits}) should be slightly modified for a two-component Galactic model]. These values lead to a set of equations, relating the parameters of the Galactic model. Breaking the degeneracy of the model with some assumptions than yields the parameters in a straightforward fashion.

\subsubsection{Results}

\begin{figure}[b]
\epsscale{1.2}
\plotone{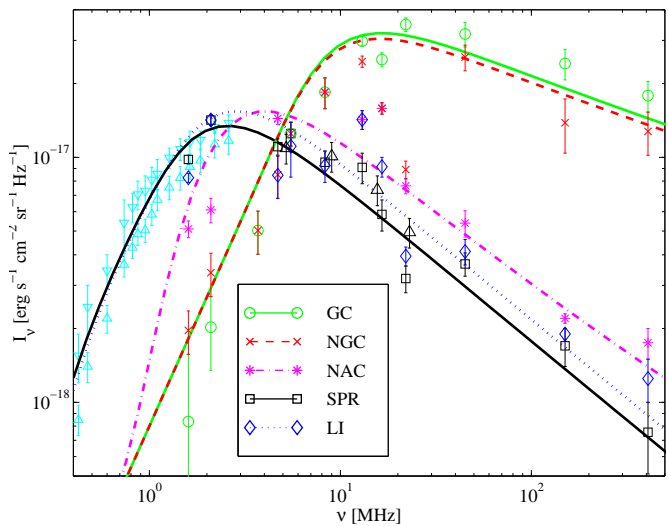}
\caption{ Demonstration of a simple model which reproduces the qualitative
features of the observed LFRB anisotropy pattern, different lines corresponding to different regions in the sky. The model consists of two Galactic disks (thin disk and thick disk) and \emph{no} extragalactic component. For model details, see \S \ref{subsec:LFRB_model}. Notations for ground-based observations (see legend) are the same as in Figure \ref{fig:LFRB_raw}, but here the lines represent the predictions of the model. The minimal and maximal brightness measured by the IMP-6 spacecraft (triangles) are for integrated large fields of view, and should thus be modelled by a frequency-dependent sky
averaging scheme.  }
\label{fig:LFRB_model}
\end{figure}

Figure \ref{fig:LFRB_model} illustrates how a simple Galactic model can account for the observed frequency-dependent anisotropy pattern of the LFRB, discussed in \S \ref{subsec:LFRB_observations}. Since the examined regions on the sky are only representative and the uncertainties are large, we can only demonstrate a qualitative agreement with the data, and argue that the model captures the important  observed features. 

The model consists of two uniform Galactic disks; a thin disk (denoted by subscript 1), containing most of the ionized gas, and a thick disk (denoted by subscript 2). The small number of sky regions examined and the uncertainty and low resolution of the data, imply that the model parameters are highly degenerate. Demanding that the $10\MHz$ synchrotron emissivity at the Galactocentric radius of the solar system ($R_s \simeq 8\kpc$) agrees with screening method measurements \cite[$j_\nu \simeq 3\times 10^{-39} \JUnits$, see] [and the references therein] {Fleishmann95}, helps break the degeneracy, yielding the best fit parameters used in Figure \ref{fig:LFRB_model}. The thin disk is found to have a radius $R_1\simeq 8.5\kpc$, a half scale height $z_1(R_s)/2 \simeq 20\pc$ at the Galactocentric radius of the solar system, synchrotron emissivity $j_1(\nu)\simeq 6.6 \times 10^{-40} (\nu/10\MHz)^{-0.25} \IUnits$, and an absorption coefficient $\alpha_1(1\MHz) \simeq 1.5\times 10^{-21} \cm^{-1}$. For the thick disk, we find $R_2 \simeq 10\kpc$, $z_2(R_s)/2 \simeq 1\kpc$, $j_2(\nu) \simeq 2.4 \times 10^{-39} (\nu/10\MHz)^{-0.65} \IUnits$ and $\alpha_2(1\MHz) \simeq 1.4\times 10^{-21} \cm^{-1}$.

The dimensions of the Galactic disks found above are in qualitative agreement with concurrent Galactic models, based on various observations that measure emission and absorption processes in the ISM, and propagation effects of radiation passing through it \cite[see the discussion of spatial models in \S \ref{subsec:LFRB_prev_models};] [] {Phillips81a, Cordes02, Keshet04a}. However, the model parameters should be considered as order of magnitude estimates only, because they depend upon the modelled regions (chosen by the availability of multi-frequency data) and on the choice of degeneracy breaking. In addition, these values represent averages over large, $\sim \kpc$ distances, where large fluctuation exist. 

For example, the emissivity of the thick disk is found to be a factor of $\sim3.5$ higher than the emissivity of the thin disk for $\nu=10\MHz$ (the two become comparable for $\nu\simeq 240\MHz$). However, this reflects the sensitivity of the high latitude observations modelled to the emissivity in the Galactocentric radius of the sun, known to be enhanced within the local arm \cite{Caswell76, Fleishmann95}, and does not imply that the emissivity is overall larger in the thick disk. Similarly, in our simple model, the value of $\alpha$ is strongly constrained by measurements in low frequencies, where the optical depth is large and the radiation observed originates from the vicinity of the sun. Hence, the modelled value of $\alpha$ reflects the state of the ISM in the solar neighborhood, and is unsurprisingly found to be similar in both disks. The absorption coefficient provides a measure of the thermodynamical state of the modelled ISM, roughly proportional to the value of the combination $n_e^{2} T^{-3/2}$ averaged over the different phases of the ISM along the line of sight, where $T$ is the temperature and $n_e$ is the electron number density. For typical values $\alpha(1\MHz) \simeq 1.5 \times 10^{-21} \cm^{-1}$ and a local density $n_e\simeq 0.05\cm^{-3}$ \cite[e.g.] [] {Reynolds91}, we find $\langle T \rangle \simeq 2500 \K$. These estimates for the average values of $J(\nu)$, $\alpha(\nu)$ and $T$ may not be extrapolated to large distances away from the solar system based on the present analysis. 

Our model is highly oversimplified, lacking for example spiral structure,
radial gradients along the Galactic plane, and north-south Galactic
asymmetry. It is important to note, however, that within observational
uncertainties, the model provides a good fit to the data without the
introduction of an additional, isotropic extragalactic component. Hence, the data does not directly imply the existence of a DERB. The frequency-dependent anisotropy pattern of the LFRB may be used in the future, not only to construct more realistic Galactic models, but also to measure the extragalactic radio background or to place stringent limits on its magnitude. For example, improved data may allow one to study the brightness of the polar regions at low optical depths, and to measure the deviation from an angular disk-like profile or from a pure power-law spectrum. With the present data uncertainties, however, we can only impose upper limits on the DERB flux which are of the same order as the measured Galactic signal.

\subsection{Previous models}
\label{subsec:LFRB_prev_models}

In the following we present the main LFRB models that have appeared in the literature. Although various models at different levels of sophistication have been proposed to account for either spectral or angular features of the observed LFRB, no comprehensive model has thus far accounted for the combined spectral and angular data. In particular, we are not aware of any model that explains the frequency-dependent anisotropy pattern of the LFRB discussed above, although the basic concept of enhanced absorption at low latitudes has been acknowledged long ago \cite[e.g.][]{Ellis82}. In general, models accounting for the spectral features observed (hereafter spectral models) yield little information about the structure of the Galaxy, but can in principle separate between a Galactic and an extragalactic component. Models accounting for the angular data (hereafter spatial models) can be used to study the structure of the Galaxy, but do not provide information regarding the extragalactic component; the latter must therefore be separately specified and subtracted from the data.

\subsubsection{Spectral Models}
\label{sec:spectral_models}

Spectral models have been used to explain the radio spectrum measured towards a given beam direction, without accounting for the anisotropy pattern. Such models are especially appropriate for low resolution LFRB observation, such as satellite measurements, which provide mostly spectral information. Spectral models often make two simplifying assumption: (i) the radio signal is composed of Galactic foreground and an extragalactic background, represented by the (oversimplified) two-component model discussed in \ref{subsec:LFRB_model} [see equation (\ref{eq:two_comp})]; and (ii) both Galactic and extragalactic components are assumed to have intrinsic power-law spectra, $j(\nu) \propto \nu^{-s_g}$ and $I_x(\nu) \propto \nu^{-s_x}$, at low frequencies (say, $\nu\la200\MHz$). With these assumptions, one may use deviations of the observed intensity from a pure power-law, in order to fit the parameters of the model. 

In this method, Clark et al. (1970) analyzed the emission from the 'north halo minimum', a region of minimal radio brightness around Galactic coordinates $l=150\de$ and $b=50\de$, which features a slight excess of emission in frequencies $2-4\MHz$. Attributing this excess to extragalactic emission yields $s_g\simeq 0.4 \pm 0.05$, $s_x\simeq 0.8 \pm 0.1$ and $I_x(10\MHz)\simeq(3.5 \pm 1.3) \times 10^{-18}\IUnits$. However, in order to fit the data, the extragalactic background must cut off sharply below $3\MHz$, presumably because of strong extragalactic absorption. Cane (1979) has analyzed the emission from the Galactic poles, and reported that although the data may be explained by a uniform Galactic disk, a better fit is obtained by including an extragalactic component.  By assuming that $s_x\simeq 0.8$, she finds $s_g\simeq 0.51\pm0.02$ and $I_x(10\MHz) \simeq(1.6 \pm 0.5) \times 10^{-18}\IUnits$. The synchrotron emissivity has been estimated by attributing the high latitude radio brightness to the integrated emissivity through a uniform disk of some assumed half scale height $z$. Thus, Cane (1979) assumed $z=500\pc$, and found $j(10\MHz)\simeq (4.9\pm 0.3) \times 10^{-39} \JUnits$.

In addition to the shortcomings of the above oversimplified model, as
discussed in \S \ref{subsec:LFRB_model}, such spectral models may in
general lead to an \emph{overestimated} extragalactic low-frequency radio
background. The reason for this is that the modelled Galactic emission
alone provides an acceptable fit to the spectral data, such that the
additional extragalactic component must be smaller than the Galactic
component at all frequencies and in all directions. Hence, the reported
extragalactic background was found to be on the order of the expected
integrated emission from discrete radio sources \cite{Cane79}, or somewhat stronger but intrinsically cutting off below $\sim 3\MHz$ \cite{Clark70}. The preceding discussion indicates, that without a \emph{significant} spectral feature, that persists over various lines of sight, and considering the present observational uncertainties and limited understanding of Galactic radio emission, spectral models can not reliably identify the extragalactic component. Spectral features observed at very low frequencies ($\nu<10\MHz$), where the spectrum of Galactic emission strongly deviates from a power-law, may in principle be used to estimate the DERB flux, but (i) with present uncertainties such estimates can only be used as upper limits; and (ii) these limits are only applicable for very low frequencies, say $\nu < 30\MHz$. Independent attempts to estimate the extragalactic emission by making assumptions regarding the uniformity of the Galactic spectral index $s_g$ \cite[e.g.][]{Bridle67}, are sensitive to this inaccurate assumption and may misinterpret an isotropic Galactic component as being extragalactic, as pointed out by Lawson et al. (1987).

\subsubsection{Spatial Models}
\label{sec:spatial_models}

More sophisticated models of Galactic radio emission \cite{Phillips81a, Beuermann85} have been constructed by analyzing detailed all-sky maps at frequencies where absorption is negligible, such as the Bonn all-sky radio continuum survey at $408\MHz$ \cite{Haslam82}. Such three-dimensional models generally identify the distribution of Galactic cosmic-ray electrons and magnetic fields, as composed of a disk or a combination of disks, with spiral structure directly evident from steps in the radio temperature profile observed along the Galactic plane \cite{Mills59}. Unfolding procedures, whereby the brightness in a given line of sight is related to the various spiral sections contributing to it, were used to calculate the locations and the relative emissivities of the Galactic spiral arms. 
While the negligible absorption considerably simplifies the analysis and allows one to reconstruct the structure of the Galaxy, it is difficult to distinguish between the Galactic signal and the extragalactic background. Spatial models thus assume (and remove) some isotropic extragalactic component. For example, a component of brightness temperature $\sim6\K$ is often uniformly subtracted from the $408\MHz$ temperature map in order to account for extragalactic emission, assumed to be composed of similar contributions from the CMB and from discrete radio sources. 

As an example, Phillips et al. (1981a) have unfolded the emission originating from the Galactic plane, finding spiral structure at distances between $3.6\kpc$ and $16-20\kpc$ from the GC with at least three spiral arms of an arm--inter-arm emissivity ratio $\ga$ 5, and placing the solar system between two major arms. 
The radial emissivity profile within a given spiral arm is degenerate in such models. Hence, a good fit to the data was found for various profiles, including an exponential profile of scale length $3.9\kpc$, provided that the Galaxy contains roughly equal magnitudes of regular (oriented along its spiral structure) and random magnetic fields.
Extending the analysis to the entire sky \cite{Phillips81b} indicates that $\sim90\%$ of the emission originates from a disk of HWHM thickness $0.46-0.85\kpc$ at the Galactocentric radius $R_s$ of the sun, and $\sim10\%$ originates from a much thicker disk, extending $\sim10\kpc$ from the Galactic plane. Both disks grow thicker at larger distances from the GC, resembling the thickening of the atomic hydrogen disk outside the solar radius \cite{Jackson74}. A good fit to the data was obtained when the thinner disk scales as $\sim \exp(-5+R/10\kpc)$. The implied emissivity at the solar neighborhood is $j(408\MHz) \simeq (0.4-1.9) \times10^{-40} \JUnits$, suggesting the presence of regular and irregular magnetic fields of magnitude $3-6\muG$ each. Extrapolating a typical emissivity $j(408\MHz)\simeq 1.5\times 10^{-40}\JUnits$ to $10\MHz$ frequencies, by assuming an effective spectral index $s_{\mbox{\tiny eff}}=0.6$ in the range $10-408\MHz$, gives $j(10\MHz)\simeq 1.4\times 10^{-39}\JUnits$. 

Clearly, such single-frequency studies are limited in their capability to reconstruct the structure of the Galaxy, as evident from the degeneracies of the model (e.g. in the radial profile) and some of its non-physical features (e.g.  the exaggerated scale height of the thick disk). By performing a detailed spatial analysis using such unfolding procedures, but for sky maps measured in more than one frequency, in particular at low frequencies where absorption is not negligible, one may construct a far better model of Galactic emission and absorption than currently available. The combined spectral and angular data will remove some of the degeneracies present in both spectral and spatial models, relax some of their simplifying assumptions, and possibly provide a direct estimate of the extragalactic component. Such an approach requires low frequency sky maps of better resolution than presently available, and is beyond the scope of this paper. However, future very low ($\nu\la10\MHz$) frequency observations, for example by the ALFA mission, may enable such an analysis.


\section{Discussion}
\label{sec:discussion}

We have studied the radio signal produced by synchrotron emission from the
strong intergalactic shocks associated with structure formation (see \S \ref{sec:IGM_model}). The analytical model of Waxman \& Loeb (2000) was generalized by adapting it for a \LCDM universe, incorporating spectral features and shock asymmetry into the model, and calibrating the free parameters of the model, as summarized in Table \ref{tab:ModelParams}. The halo parameters $f_{acc}$, $f_T$, and $f_{r}$ were calibrated using a hydrodynamical cosmological simulation, by demanding that the model agrees with the simulation on the average baryon temperature, the average mass consumption rate by strong shocks, and the typical size of regions where the thermal energy is injected by shocks. After calibrating the model with these essentially global features, it yields radio (see Figures \ref{fig:calculated_spectrum}, \ref{fig:calculated_correlation} and \ref{fig:synch_Cl}) and \gama-ray \cite[see] []{Keshet03} signals which are in good agreement with the signals extracted independently from the simulation. Although the parameter calibration scheme is inaccurate and the weak redshift dependence of the parameters needs yet to be determined by a more careful analysis, this agreement suggests that the calibration procedure is sensible. The localized nature of regions where most of the thermal energy is injected \cite[e.g. at the intersections of X-ray cluster accretion shocks with galaxy filaments, channelling gas into clusters, see] [] {Keshet03}, enhances the synchrotron luminosity of a halo while leaving its inverse-Compton luminosity unchanged, because of the enhanced magnetic energy density. The radio luminosity-temperature relation according to the calibrated model was shown to be in qualitative agreement with observations of cluster radio halos (see Figure \ref{fig:clusters}). 

We have examined the observational consequences of the predicted radio signal, for present-day and for future radio telescopes (see \S \ref{sec:feasibility}). Our main results are illustrated in Figures \ref{fig:LFRB_sky} and \ref{fig:Sky_Cl} (for the calibrated model parameters discussed in \S \ref{subsec:model_LCDM_sim} and summarized in Table \ref{tab:ModelParams}). Figure \ref{fig:LFRB_sky} shows the contribution of various signals to the radio sky, suggesting that emission from intergalactic shocks contributes up to a few tens of percent of the extragalactic radio background below $500\MHz$ (see also \S \ref{sec:LFRB}). Figure \ref{fig:Sky_Cl} depicts the angular power spectrum of various signals on an angular scale of $\sim0\de.5$, along with the sensitivities and the angular resolutions of the telescopes. The figure demonstrates that the designs of next generation radio telescopes such as the LOFAR and the SKA, are more than sufficient for a detection of the angular fluctuations introduced by intergalactic shocks. In fact, even present-day high-resolution radio telescopes (such as the VLA in its maximum sensitivity configuration), are potentially sensitive to the predicted signal.

Foreground fluctuations in the synchrotron emission of our Galaxy limit a
positive detection of the signal to sub-degree scales and to high Galactic
latitudes. Confusion with discrete radio sources requires that the brightest sources be modelled and removed. With a feasible point source cut ($100\mJy$ at $1.4 \GHz$ or equivalently $3\Jy$ at $150\MHz$), the intergalactic signal dominates over all foreground and background signals, at angular scales $10\arcmin \la \theta \la 1\de$ and frequencies $\nu<10\GHz$ (assuming $\xi_e \xi_B \ga 3\times 10^{-4}$, see \S \ref{subsec:model_efficiency} and the discussion below). For the above cut, the signal is of the same magnitude as the integrated emission from discrete sources at $\sim1\arcmin$ scales. Attempts to detect the signal at $\la1\arcmin$ scales are confusion limited by discrete sources, demanding increasingly more ambitious point source removal schemes. In addition to the labor involved, noise is always introduced when removing discrete sources, because of source model uncertainties. The spectrum of the signal suggests that it is most pronounced at frequencies around $\sim 100\MHz$, planned to be covered by both the LOFAR and the SKA.

The calculated level of angular fluctuations in the radio emission from
intergalactic shocks is sensitive to uncertainties in the calibration of
the model parameters. The logarithmic contribution to the variance (shown
in Figure \ref{fig:Sky_Cl}) scales roughly according to \beq
\delta I_l(l;\,\xi_e,\xi_B,f_{acc},f_T,\widetilde{f}_{r}) = \xi_e \xi_B {f_{acc} f_T^2 \over \widetilde{f}_{r}^3} \, \delta I_l (\widetilde{f}_{r}l; \,1,1,1,1,1) \fin \eeq 
The radio signal calculated from the \LCDM simulation \cite{Keshet04b}, however, depends only on the energy fractions $\xi_e$ and $\xi_B$, through the combination $\xi_e \xi_B$. The results of the simulation, as shown in Figures \ref{fig:calculated_spectrum}, \ref{fig:calculated_correlation}, and \ref{fig:synch_Cl}-\ref{fig:Sky_Cl}, suggest that $\delta I_l$ was in fact slightly \emph{underestimated} by our parameter calibration scheme. This implies, for example, that intergalactic shocks introduce intensity fluctuations of magnitude $\delta I_l \ga 8\times 10^{-19} (\xi_e \, \xi_B / 5\times 10^{-4}) (\nu/100\MHz)^{-1} \IUnits\!\!$ on multipoles $400\la l \la 2000$, corresponding to temperature fluctuations $\delta T_l \ga 260 (\xi_e \, \xi_B / 5\times 10^{-4}) (\nu/100\MHz)^{-3} \K$. 

Our model implies that emission from intergalactic shocks on angular scales $\sim 0\de.5$ dominates the sky at $<500\MHz$ frequencies if $\xi_e \xi_B\ga 10^{-4}$, and is even dominant at frequencies as high as $10\GHz$ if $\xi_e \xi_B\ga 3 \times 10^{-4}$. The simulation suggests that the signal is stronger than predicted by the (calibrated) model, in particular on small angular scales. Therefore, according to the simulation, emission from intergalactic shocks will dominate the sky at $<0\de.5$ scales and $<500\MHz$ frequencies, even if $\xi_e \xi_B \simeq 5\times 10^{-5}$. We have used observations of SNR shocks and of magnetic fields in the halos of galaxy clusters, in order to show that for strong intergalactic shocks $\xi_e \xi_B \simeq 5 \times 10^{-4}$, unlikely to be smaller than this value by more than a factor of $\sim 4$. For this purpose, in \S \ref{subsec:model_efficiency} we have used dimensional analysis arguments to show that the physics of strong intergalactic shocks is essentially identical to the physics of strong SNR shocks, both of comparable velocities $v\simeq 10^3\km \se^{-1}$, provided that an appropriate re-scaling of time is carried out \cite[see also] [] {Keshet03}. 
We have assumed that strong shocks accelerate electrons to a power law distribution of index $p=2$ (equal energy per logarithmic interval of electron energy). Such a distribution is inferred from SNR observations and agrees with linear models for diffusive shock acceleration, although non-linear models suggest deviations from a pure power-law. 

As noted by Waxman \& Loeb (2000), emission from intergalactic shocks may have already been detected by CMB anisotropy studies at frequencies $\la 10\GHz$ and angular scales smaller than a degree. In particular, telescopes operating at frequencies of a few GHz such as ACSA, Ryle and the VLA, may have detected the signal at arcminute scales. High resolution $\sim 1\GHz$ radio surveys may have also detected the signal by now at an identifiable level. Such surveys are generally available at low Galactic latitudes, where contamination by Galactic foreground fluctuations is severe. Nevertheless, analysis of 'quiet regions' observed off the Galactic plane, including a careful removal of discrete sources, may yield the desired signal. It will probably be easiest to identify the signal at $1\arcmin-10\arcmin$ scales, by modelling discrete sources and cross correlating the maps with known tracers of large-scale structure. Detection of emission from intergalactic shocks is not unrealistic even in case the signal has been mildly overestimated. A signal $\delta I_l$ lower than calculated in \S\ref{sec:IGM_model} by a factor of a few, may still be identified at $\sim 1\arcmin-10\arcmin$ scales, if faint point sources are modelled and removed. If the signal was overestimated by $\sim 2$ orders of magnitude, it may still be detectable by next generation radio telescopes, by means of cross-correlation with known tracers of large-scale structure such as galaxy counts, or, in the future, with \gama-ray emission from intergalactic shocks.  

As mentioned in \S \ref{sec:Introduction}, future detection of radio emission from intergalactic shocks will have important implications on our understanding of cosmology and astrophysics. Detection of the signal will provide the first identification of intergalactic shocks, revealing the underlying cosmological flows and providing a test for structure formation models. The signal, in particular when combined with \gama-ray detection, will provide a measure of the unknown magnetic fields in the intergalactic medium. 
Although non-trivial for interpretation, such a measure of the magnetic field may provide insight into the unknown processes leading to IGM magnetization. The signal may also confirm the existence of the undetected warm-hot intergalactic medium, and provide a tracer for its distribution. 

Synchrotron emission from intergalactic shocks is correlated with the large-scale structure of the low-redshift ($z<1$) universe, tracing young galaxy clusters and filaments. The signal could thus account for some observed features of the radio signature of galaxy clusters, namely radio halos and radio relics. Extended accretion shocks could contribute to radio halos \cite[e.g. to the large radio halo of the Coma super-cluster, see][and the references therein]{Thierbach03}, whereas localized shocks (e.g. where galaxy filaments channel large amounts of gas into the cluster regions) may be responsible for some radio relics observed at the outskirts of clusters [such as the prototype relic found in the Coma super-cluster, 1253+275 \cite{Ensslin98}, and the large-scale radio arcs observed in clusters A3667 \cite{Rottgering97} and A3376 \cite{Bagchi02}], and perhaps also for some of the anomalous features observed in radio halos \cite[e.g. in the unrelaxed clusters described by] [] {Govoni04}. 

Radio emission from intergalactic shocks is an important source of
contamination for other radio signals, such as the low frequency CMB,
Galactic synchrotron fluctuations on sub-degree scales, and competing
extragalactic radio signals such as bremsstrahlung from \Lya clouds, and 21
cm tomography (Zaldarriaga et al. 2003; Morales \& Hewitt 2003; Loeb \& Zaldarriaga 2003; Gnedin \& Shaver 2003). 
For example, the integrated emission from intergalactic shocks introduces distortions in the low-frequency average brightness temperature, at the level of $\delta T \simeq 1.6 \,(\nu / 3 \GHz)^{-3} \mK$. Efforts to measure distortions in the CMB spectrum at low frequencies, e.g. in the next mission of the Absolute Radiometer for Cosmology, Astrophysics and Diffuse Emission (ARCADE\footnote{See http://arcade.gsfc.nasa.gov/arcade/index.html}), may thus be sensitive to the signal at the low frequency bands \cite[$\sim 3\GHz$, see] [] {Fixen04,Kogut04}. 
We have shown that radio emission from intergalactic shocks constitutes a
significant fraction of the extragalactic radio background below
$500\MHz$. This component should thus be taken into account when evaluating
the propagation of ultra-high energy photons with energies above
$10^{19}\eV$, because at these energies the effect of the radio background on the transparency of the universe is stronger than the effect of the CMB, and may thus be important in models for the origin of ultra-high energy cosmic-rays.

We have analyzed observations of the diffuse low frequency radio background below $500\MHz$ (see \S \ref{sec:LFRB}), and highlighted the frequency-dependent anisotropy pattern observed in frequencies $1\MHz \la \nu \la 10\MHz$ (see Figure \ref{fig:LFRB_raw}). We presented a simple Galactic model, consisting of a thin disk, containing most of the ionized gas, and a thick disk. This model was shown to provide a good fit to the data (see Figure \ref{fig:LFRB_model}), considering the observational uncertainties. The model thus enabled us to (i) assess the feasibility of directly inferring the flux of the diffuse extragalactic radio background from presently available observations; and (ii) demonstrate how an elaborate Galactic model may be constructed in a straightforward fashion, using the anisotropy pattern observed in very low frequencies. 

The agreement of the Galactic model with observations, and the large uncertainties and poor resolution of present low frequency observations, preclude a direct identification of the diffuse extragalactic radio background. At best, an upper limit on the extragalactic component may be imposed, at very low ($\la 30$ MHz) frequencies where free-free absorption in the Galaxy becomes important. Such a flux constraint, a factor of a few lower than the Galactic foreground, is itself higher than the calculated signal from discrete sources and from intergalactic shocks by a factor of a few. This implies that (i) the true extragalactic background in $\nu\simeq 10\MHz$ frequencies is probably smaller than the Galactic foreground by a factor of a few; and (ii) emission from intergalactic shocks constitutes a significant fraction of the total extragalactic signal, measuring up to a few tens of percent.

We have reviewed previous models of the low frequency radio background, and demonstrated that previous claims for direct identification of the diffuse extragalactic component are based on insignificant, probably Galactic, spectral features. We pointed out that previous models fail to account for the combined spectral and angular data at very low ($1\MHz<\nu<10\MHz$) frequencies, where absorption is non-negligible and the anisotropy pattern of the sky is highly frequency dependent. Modelling the combined data can be used to disentangle the distributions of Galactic cosmic-rays, ionized gas and magnetic fields, and lead to an elaborate three-dimensional model of the Galaxy. Our model is too simplistic, and the data uncertainties too large, to provide reliable estimates of the physical properties of our Galaxy, other than confirming its double-disk structure and the presence of a $\sim 1\kpc$ thick disk of cosmic ray electrons and magnetic fields, and imposing constraints on free-free absorption in the solar neighborhood. Future high resolution observations at very low frequencies, produced for example by the ALFA space mission, could thus provide valuable information regarding the structure and the composition of the Milky Way.

\acknowledgments 
This work was supported in part by grants from NSF (AST-0204514 and AST-0071019) and NASA (NAG-13292), by a MINERVA grant and by an AEC grant. We thank the Institute for Advanced Study for its kind hospitality during a period when this project was initiated. We thank the anonymous referee for useful comments.



\end{document}